\documentclass[useAMS,usenatbib]{mn2e}
\usepackage{graphicx}
\usepackage{lscape}
\usepackage{ulem}
\def \hMpc{h^{-1}{\rm\ Mpc}}
\def\kms{km~s$^{-1}$}

\def\lya{Ly$\alpha$}

\def\hkpc{$h_{70}^{-1}$ kpc}

\newcommand{\Msun}{M_{\odot}}

\title[Damped Lyman alpha absorption in a binary QSO]
{Coincident, 100 kpc-scale damped Lyman alpha absorption towards a binary 
QSO: how large are galaxies at $z \sim 3$?}
\author[S. L. Ellison et al.]
{Sara L. Ellison$^{1,}$\thanks{Email: sarae@uvic.ca},
Joseph F. Hennawi$^{2,3}$,
Crystal L. Martin$^{4,5}$,
Jesper Sommer-Larsen$^{6,7}$.
\\
$^1$Dept. Physics \& Astronomy, University of Victoria, 3800 Finnerty Rd, 
Victoria, BC, V8P 1A1, Canada\\
$^2$Dept. Physics \& Astronomy, University of California, Berkeley, USA\\
$^3$Hubble Fellow\\
$^4$Dept. Physics \& Astronomy, University of California, Santa Barbara, USA\\
$^5$Packard Fellow\\
$^6$Dark Cosmology Centre, Niels Bohr Institute, Juliane Maries Vej 30, 
   DK-2100 Copenhagen {\O}, Denmark\\
$^7$Institute of Astronomy, University of Tokyo, Osawa 2-21-1,
   Mitaka, Tokyo, 181-0015, Japan
}

\begin{document}
\maketitle

\begin{abstract}

We report coincident damped Lyman alpha (DLA) and sub-DLA absorption at
$z_{\rm abs} = 2.66$ and $z_{\rm abs} = 2.94$ towards the $z \sim 3$,
13.8 arcsecond separation binary quasar SDSS 1116+4118 AB.  At the redshifts of
the absorbers, this angular separation corresponds to a 
proper transverse separation of $\sim$ 110 \hkpc.  A third absorber,
a sub-DLA at $z_{\rm abs} = 2.47$, is detected towards SDSS 1116+4118 B,
but no corresponding high column density absorber is present
towards SDSS 1116+4118 A.  We use high resolution galaxy simulations 
and a clustering analysis to
interpret the coincident absorption and its implications for galaxy
structure at $z \sim $3.  We conclude that the common absorption in the
two lines of sight is unlikely to arise from a single galaxy, or a galaxy
plus satellite system, and is more feasibly explained by a group of
two or more galaxies with separations $\sim$ 100 kpc.  The impact
of these findings on single line of sight observations is also discussed;
we show that abundances of DLAs may be affected by up to a few tenths
of a dex by line of sight DLA blending. From a Keck ESI spectrum of the two
quasars, we measure metal column densities for all five absorbers
and determine abundances for the three absorbers with log N(HI) $>$ 20.
For the two highest N(HI) absorbers, we determine high levels of
metal enrichment, corresponding to 1/3 and 1/5 $Z_{\odot}$.  
These metallicities
are amongst the highest measured for DLAs at any redshift and
are consistent with values measured in Lyman break galaxies at
$2 < z < 3$.  For the DLA at  $z_{\rm abs} = 2.94$ we also 
infer an approximately solar ratio of $\alpha$-to-Fe peak elements
from [S/Zn] = +0.05,
and measure an upper limit for the molecular fraction in this particular
line of sight of log $f($H$_2)< -5.5$.

\end{abstract}


\section{Introduction}

For the last decade, a simple schematic view of the relative sizes
of QSO absorption systems has been built on the coupling of
absorbers with galaxies of varying luminosities and impact parameters
and the consideration of luminosity functions and
number densities (e.g. Steidel 1993; Steidel 1995; Lanzetta 1993).
Although a galaxy's gas cross-section depends on its individual
properties (such as mass and luminosity), the topology of a given
galaxy is usually considered to be hierarchical and based on
column density (e.g. Steidel 1993; Churchill, Kacprzak \&
Steidel 2005). In this picture, the damped Lyman alpha
(DLA) systems represent a relatively small fraction of a galaxy's
gas cross-section, associated with its inner $\sim$ 10--20 kpc.  The Mg~II
bearing gas, associated with Lyman limit systems, occupies a somewhat
larger halo; in the original Steidel (1995) picture this halo 
was roughly spherical, had
a covering factor of approximately unity and a radius $\sim$ 40 kpc
for an $L^{\star}$ galaxy.  The largest absorption cross-section was
associated with C~IV absorbing gas and extended out to distances on
the order of 100 kpc.  Although attractive in its simplicity, this
picture has recently undergone significant re-evaluation.  
For example, although the
idea of large C~IV halos has been vindicated by observations of absorbers
near to Lyman break galaxies (LBGs, Adelberger et al. 2003, 2005b), it has 
been argued that some CIV absorbers may be associated with a more
diffuse component, possibly the intergalactic medium (Pieri,
Schaye \& Aguirre 2006).  Structure has also been inferred in the Mg~II
population.  Ellison et al. (2004) argued that variations in Mg~II
equivalent widths (EWs) on kpc-scales seen in spatially resolved
lensed QSO images suggest individual Mg~II `clouds' that are an
order of magnitude smaller than the halo sizes found by previous 
galaxy surveys.
Even the sizes of the Mg~II halos are currently being re-assessed;
it has been suggested by Churchill, Kacprzak \& Steidel (2005) 
that the original
survey strategies may have led to an under-estimate of the extent
of the absorbing gas.  

In light of these recent re-evaluations, it may be surprising 
how little progress we have made in determining the sizes of DLAs.
These are probably the best studied of the quasar absorption line
menagerie, partially because of the numerous chemical elements that can be
used to infer their star formation histories (e.g. Dessauges-Zavadsky
et al. 2004).   Unlike the case for C~IV and Mg~II absorbers, there is
scant data from which we can directly infer DLA sizes.  The two 
main techniques that have previously been used for this estimation, 
namely the association of individual galaxies with absorbers and the
application of lensed QSOs, are not yet on a solid statistical footing.  
The former of
these techniques requires a relatively large sample of DLAs with known
galaxy counterparts.  Although Chen \& Lanzetta (2003) attempted this
with a sample of 6 galaxies, the statistics are poor and only
available for $z < 1$, whereas the vast majority of known DLAs are at $z > 2$
(e.g. Prochaska, Herbert-Fort \& Wolfe 2005).  At higher redshifts,
only M\o ller, Fynbo \& Fall (2004) and Weatherley et al. (2005)
have made direct detections of DLAs and find impact parameters
in the range 2 -- 25 \hkpc\ for 3 DLAs.  Measurements
of DLA sizes from lensed QSOs are currently limited by the very
small number (4) of DLAs that have been detected in lensed
sightlines and by the small transverse scales that they probe.  
Two out of the four cases (Churchill et al. 2003; Kobayashi et al.
2002) probe very small scales ($<250$ pc), leaving only two measurements
on kpc scales ($d \sim 5$ kpc by Lopez et al. 2005 and $d \sim$ 10 kpc
by Smette et al. 1995).  A few indirect limits of DLA sizes
also exist, for example lower limits based on extended background
radio emission (e.g. Foltz et al. 1988; Briggs et al. 1989) or
on the unique case of transverse Ly$\alpha$ fluorescence discovered
by Adelberger et al. (2006).  It is also possible to  estimate 
gas cross-sections by combining the DLA number density with
a Holmberg relation between the radius of a disk and the galaxy
luminosity (e.g. Fynbo et al. 1999).
The disadvantage of this approach is that it assumes a specific 
and fixed geometry, although like other methods, it gives typical
disk sizes up to $\sim$ 30 kpc.
Apart from this handful of direct and indirect constraints, 
the scenario in which DLAs have 
cross-sections of a few tens of kpc has been largely untested.

In this paper, we present observations of a close binary QSO,
SDSS 1116+4118 AB, hereafter QSO A/B, with an angular separation 
between the two components (both at $z \sim 3$) of 13.8 arcseconds, 
see Table \ref{qso} and Figure \ref{sdss}.  QSO B
exhibits three high column density, intervening absorbers, two of
which are also detected in QSO A, even though the proper separation
at the redshift of the absorbers is more than 100 \hkpc.  In order
to interpret this result, we use high resolution smoothed particle 
hydrodynamic (SPH) simulations of two galaxies at $z = 2.3, 3.0, 3.6$ to
assess whether a structure as large as 100 \hkpc\ is likely to be a single
galaxy, a satellite system or a group of galaxies and  discuss the 
implications of these possibilities.

We adopt a cosmology of $\Omega_{\Lambda}$ = 0.7, $\Omega_M$ = 0.3
and H$_0$ = 70 km/s/Mpc.  In this cosmology, redshifts of 2.47, 2.66 and 
2.94 (the redshifts of the absorbers studied in this paper) 1 arcsecond
in the transverse direction corresponds to proper linear distances
of 8.09, 7.96 and 7.75 \hkpc\ respectively.

\section{Observations and Data Analysis}

\begin{figure}
\centerline{\rotatebox{0}{\resizebox{8cm}{!}
{\includegraphics{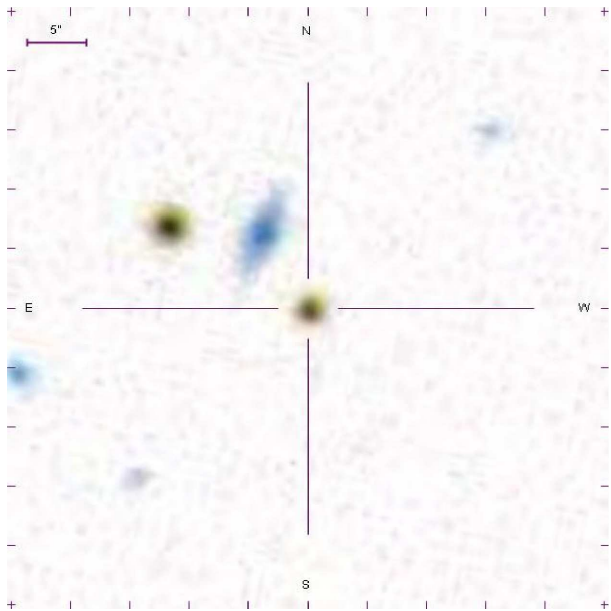}}}}
\caption{\label{sdss} SDSS image of the binary QSO SDSS1116+4118 AB.
QSO B is at the centre of the image and QSO A is offset by 13.8
arcseconds to the north-east. The galaxy located between the
two QSO images does not have a spectroscopic redshift, but
Sanchez-Alvaro \& Rodriguez-Calonge (2007) give a photometric
redshift of $z=0.25$.}
\end{figure}

Although the Sloan Digital Sky Survey (SDSS) spectroscopic quasar
survey has provided the largest sample ($\sim 10^5$) of quasars in
existence, it selects \textit{against} close quasar pairs due to the
finite size of optical fibres in the multi-object spectrograph. This
fibre collision limit implies that only one member of a pair
with $\Delta \theta < 1\arcmin$ will make it into the quasar catalog.
Hennawi et al. (2006a) selected a large sample of candidate close companions
around the SDSS quasars using photometric redshift techniques,
and spectroscopically confirmed them to be quasar pairs
with follow-up observations at low spectral resolution.  A small
number of QSO pairs can also be discovered when there is plate overlap
for a given field.  In this case, fibres can be placed on very
closely spaced objects in two different plates.
The QSO pair SDSS 1116+4118 AB falls into this latter category;
an SDSS postage stamp of the field is shown in Figure
\ref{sdss} and shows both QSOs as well as a foreground galaxy.
For a sample of confirmed spectroscopically
confirmed QSO pairs, we are currently undertaking
higher resolution spectra in order to study the transverse
absorption properties of intervening galaxies and the intergalactic
medium.  The data presented here were obtained as part of that
systematic study and they represent the only pair for which
we have currently detected a DLA in either line of sight
(many of the spectra do not have large Ly$\alpha$ forest
coverage).

Sanchez-Alvaro \& Rodriguez-Calonge (2007) have recently
proposed SDSS 1116+4118 AB to be a wide separation lens, with the
foreground galaxy (whose photometric redshift is given as $z = 0.25$)
as the lensing mass.  However, we consider this unlikely due
to a) the improved redshift determinations of the two QSOs
(Table \ref{qso}) which are significantly different: the CIV
emission lines are offset by 36 \AA; b) the
considerable differences between QSO A and B in our ESI spectra, 
both in absorption
and emission characteristics; c) the extremely low likelihood
that a single galaxy would produce such a wide separation image
(the implied mass-to-light ratio is $>$100 at the Einstein radius);
d) lack of counter images on north side of the lens galaxy
(the offset lens should produce a quadruple image); e) high
redshift ($z>2.5$) C~IV absorbers show large
fractional equivalent width differences ($>>$ 50\%) between QSO A
and B, inconsistent with lensing-predicted line of sight separations 
of $\le$ 1.5 \hkpc\ (e.g. Lopez et al. 2000; Rauch et al. 2001a,b;
Ellison et al. 2004) and f) 
the inconsistency of QSO colours, e.g. 
$g-i$=0.61$\pm$0.02 and 0.45$\pm$0.02 for A and B respectively.
Only the presence of a massive cluster would cause such wide separation
multiple images, such as the recently discovered 14 arcsecond
separation quadruply imaged SDSS 1004+4112 (Inada et al. 2003)
where the estimated cluster mass is $M \sim 10^{14} M_{\odot}$
(Oguri et al. 2004).  No such cluster is seen in the SDSS
images of SDSS 1116+4118, or in deeper MMT images (A. Marble,
private communication).  We therefore assume that SDSS 1116+4118 AB
is a projected pair of QSOs.

\begin{center}
\begin{table*}
\caption{Target Properties}
\begin{tabular}{lcccc}
\hline
QSO  & RA (J2000) & Dec (J2000) & $i$ mag & $z_{\rm em}$ \\
\hline
SDSS 1116+4118 A & 11 16 11.7 & 41 18 21.5 & 17.97 & 2.982$\pm$0.007 \\
SDSS 1116+4118 B & 11 16 10.7 & 41 18 14.4 & 19.00 & 3.007$\pm$0.007 \\
\hline 
\end{tabular}\label{qso}
\end{table*}
\end{center}

\subsection{Data Acquisition and Reduction}

On March 3 2006 we obtained 3300 seconds of integration in two exposures
on QSO A/B using
the Echellette Spectrograph and Imager (ESI, Sheinis et al. 2002)
on the Keck telescope.  Since the entrance slit of the
spectrograph is 20 arcseconds long,  the position angle was
chosen so that
both QSOs A and B were covered (see Figure \ref{sdss}).  
The observing conditions were
relatively poor, with high humidity and seeing typically 1.3 
arcseconds.  The data quality were further degraded by broad
absorption patterns across many echelle orders due to
condensation on the dewar window caused by observing in high
humidity.  Nonetheless, with a 1 arcsecond slit width and $1 \times 1$
binning the S/N ratios per pixel were $\sim$ 50 in QSO A and 30
in QSO B at 6000 \AA.

The data were reduced using a customized version of ESIRedux\footnote{
http://www2.keck.hawaii.edu/inst/esi/ESIRedux/index.html} which
was adapted to deal with multiple objects on the slit.  
Extracted spectra were calibrated to a vacuum heliocentric
wavelength scale and determined to have FWHM resolutions of
$\sim$ 60 km/s ($R \sim 5000$).  The
spectra from the two individual exposures were combined by
weighting according to S/N.  The QSO continuum (including the
broad absorption features induced by the residue on the CCD)
was estimated with the Starlink software Dipso\footnote{
http://star-www.rl.ac.uk/star/dvi/sun50.htx/sun50.html}
by fitting a cubic spline polynomial through unabsorbed regions 
of flux.  To test the effect of the CCD residue, we measure the
equivalent widths of unsaturated absorption lines that fall
in this region of the spectrum.  In theory, dividing the
spectrum by the continuum function should not alter the 
equivalents widths, if the fitting is accurate.  We find
equivalent widths that agree to within less than a few percent,
indicating that the broad absorption does not affect our
line measurements.

We identified three high N(HI) absorbers present in at least one
of the two lines of sight at redshifts $z_{\rm abs}$ =
2.47 (B only), 2.66 (A and B) and 2.94 (A and B).

\subsection{Column Density Determination}

\begin{figure}
\centerline{\rotatebox{0}{\resizebox{8cm}{!}
{\includegraphics{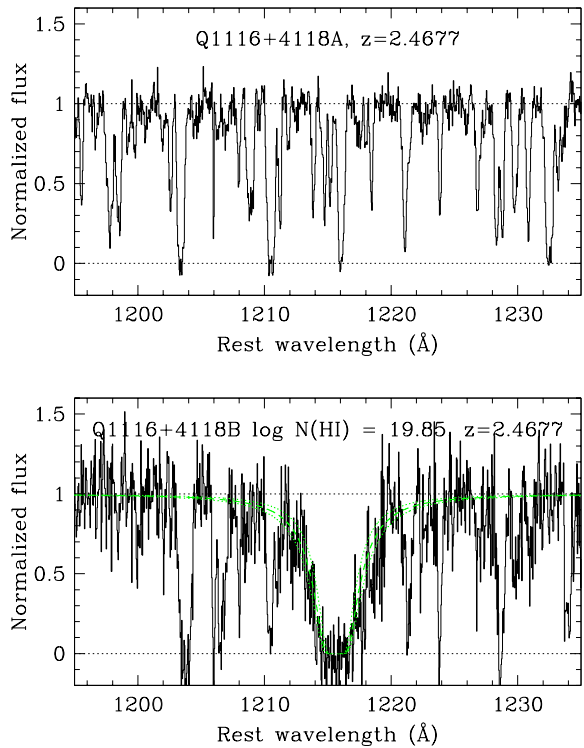}}}}
\caption{\label{hi_fit1} Ly$\alpha$ fit (dashed profile) and error
(dotted profiles) to the $z_{\rm abs} \sim 2.47$
sub-DLA seen towards QSO B (bottom panel), but not towards QSO A
(top panel). The transverse separation of the two sightlines at this redshift
is 112 \hkpc.}
\end{figure}

\begin{figure}
\centerline{\rotatebox{0}{\resizebox{8cm}{!}
{\includegraphics{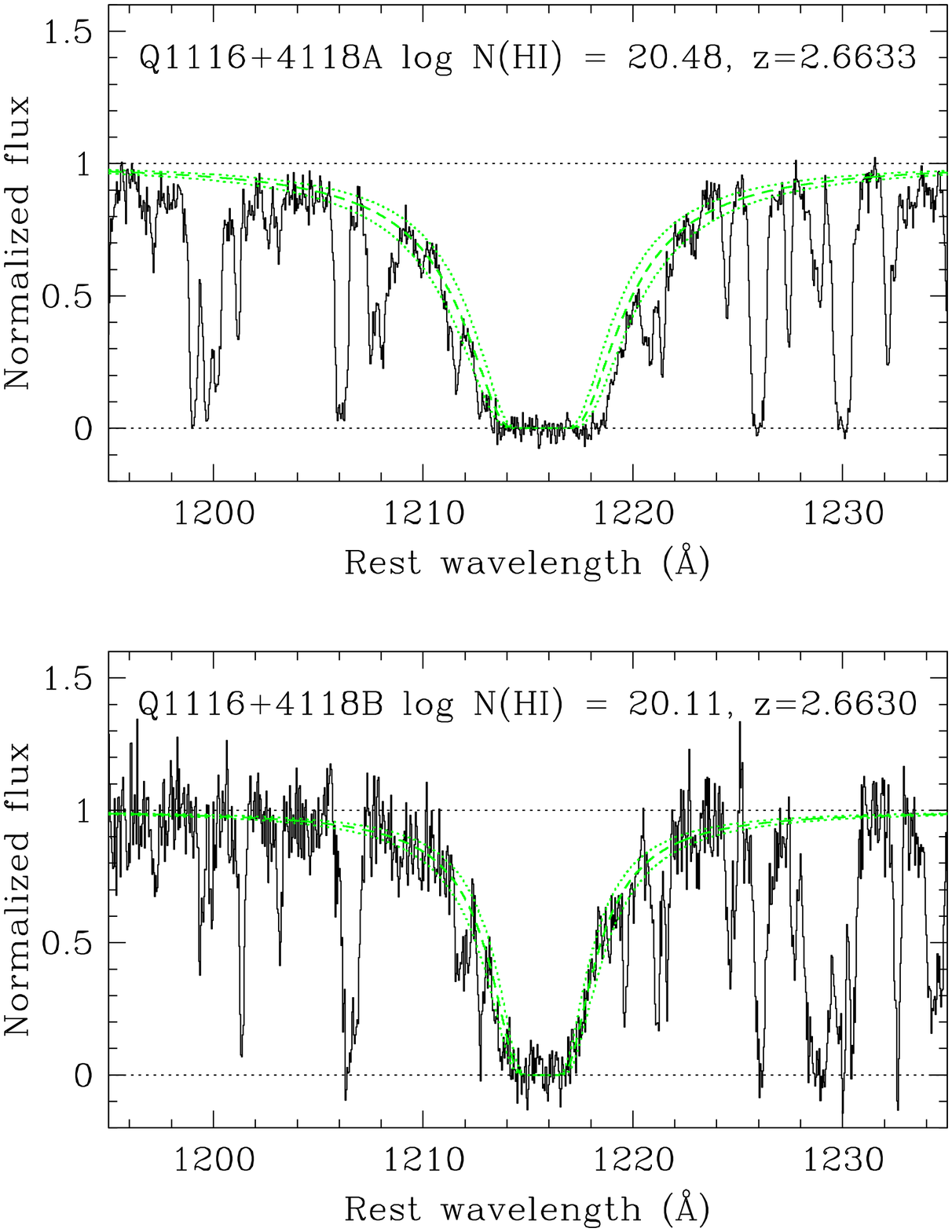}}}}
\caption{\label{hi_fit2} Ly$\alpha$ fit (dashed profile) and error
(dotted profiles) to the $z_{\rm abs} \sim 2.66$
DLA/sub-DLA seen towards QSO A (top panel) and QSO B (bottom panel).
The transverse separation of the two sightlines at this redshift
is 110 \hkpc. }
\end{figure}

\begin{figure}
\centerline{\rotatebox{0}{\resizebox{8cm}{!}
{\includegraphics{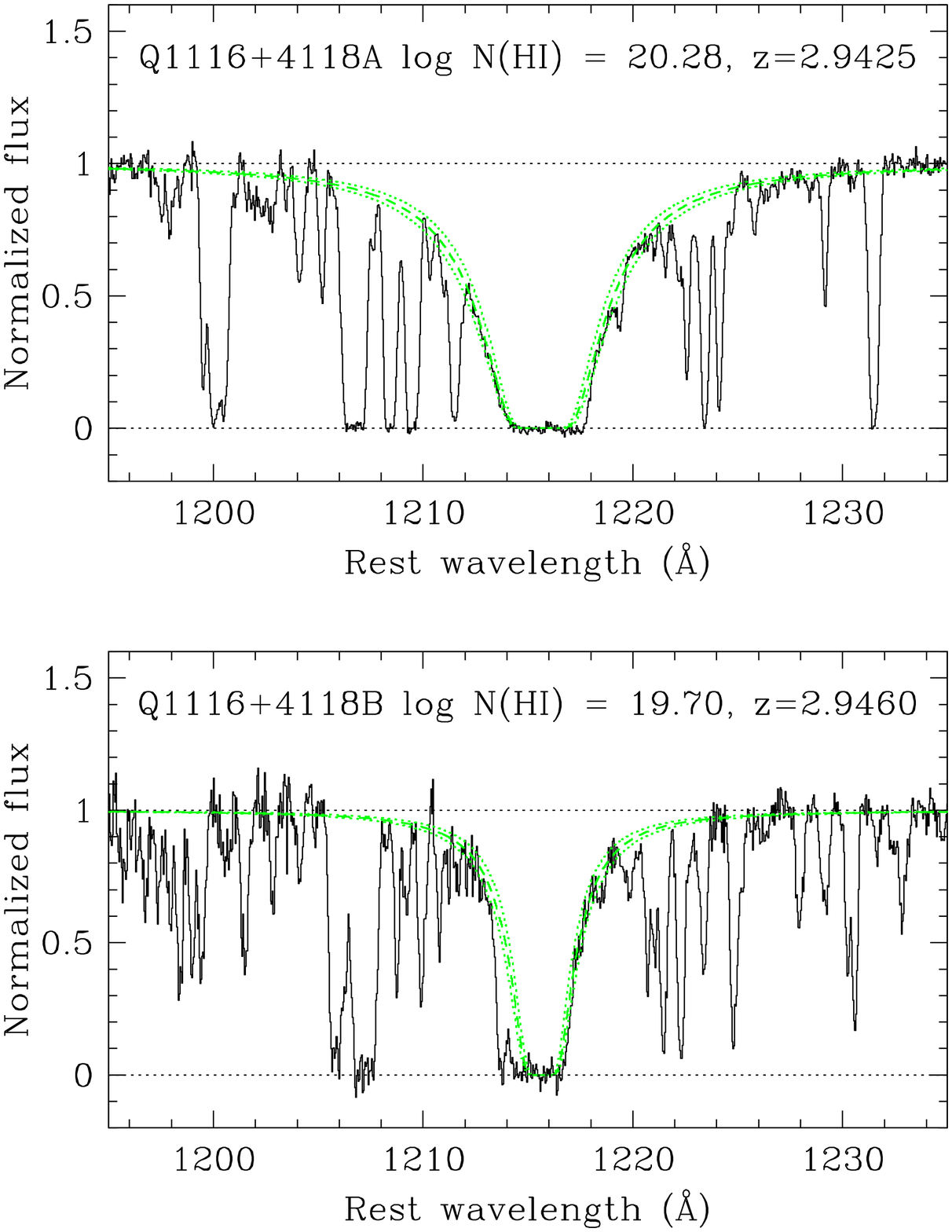}}}}
\caption{\label{hi_fit3} Ly$\alpha$ fit (dashed profile) and error
(dotted profiles) to the $z_{\rm abs} \sim 2.94$
DLA/sub-DLA seen towards QSO A (top panel) and QSO B (bottom panel).
The transverse separation of the two sightlines at this redshift
is 107 \hkpc. }
\end{figure}

\begin{figure}
\centerline{\rotatebox{0}{\resizebox{8cm}{!}
{\includegraphics{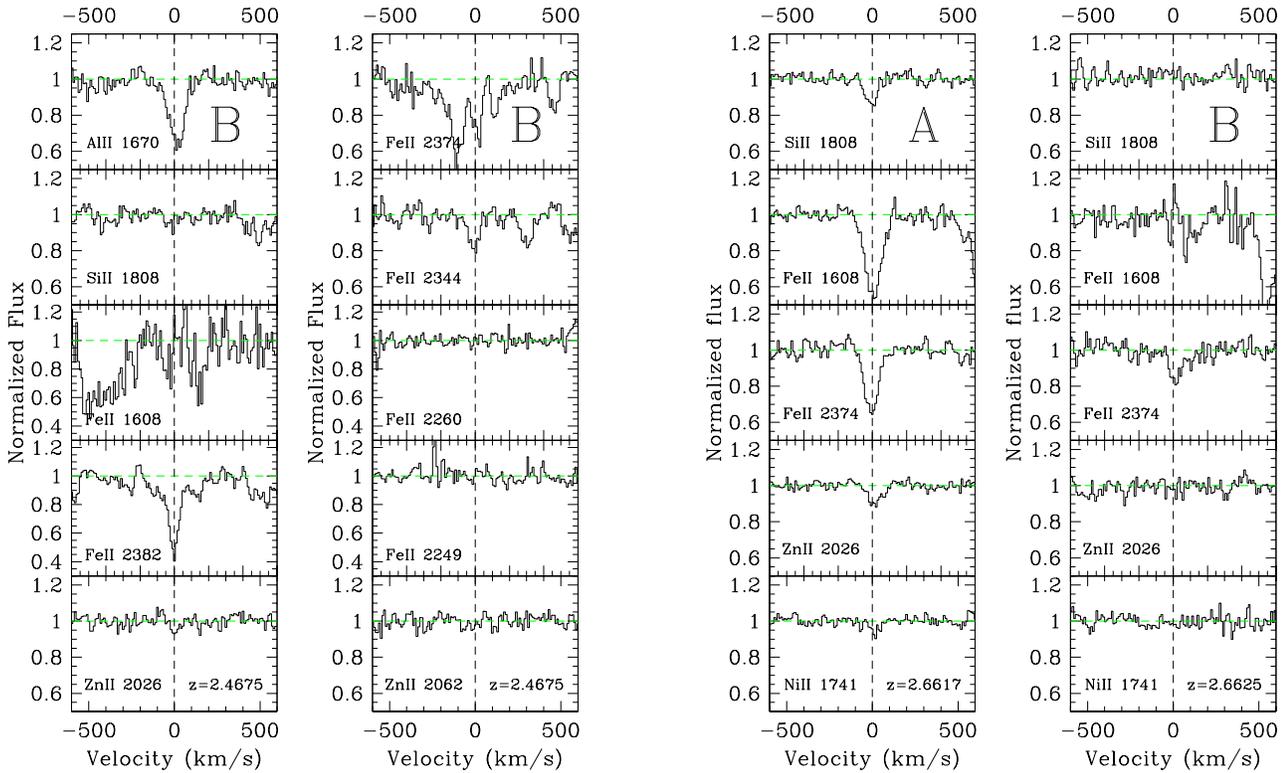}}}}
\caption{\label{met1} Metal lines for the sub-DLA detected
towards  QSO B  at $z_{\rm abs} \sim 2.4677$.  }
\end{figure}

\begin{figure}
\centerline{\rotatebox{0}{\resizebox{8cm}{!}
{\includegraphics{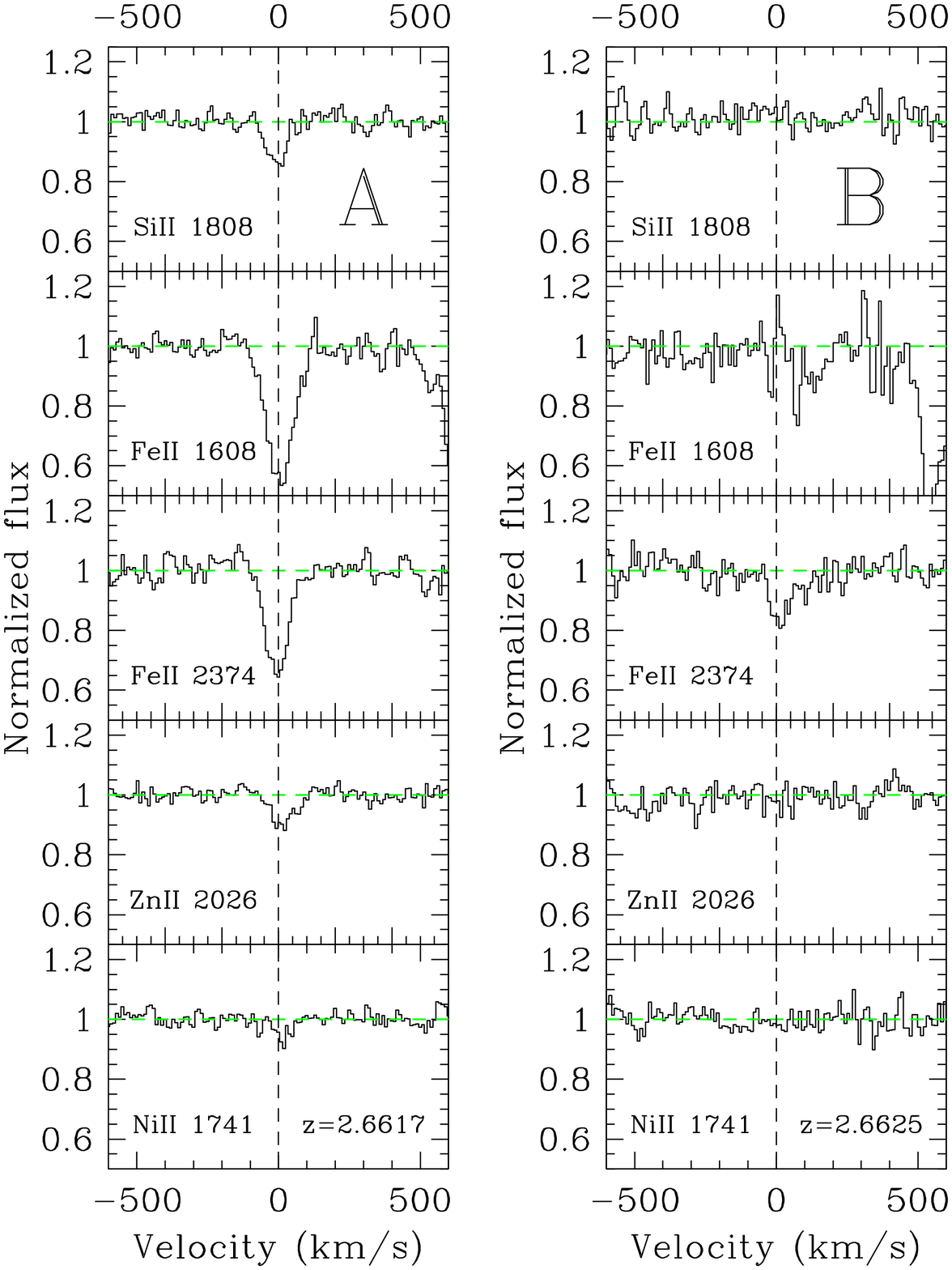}}}}
\caption{\label{met2} A selection of detected metals for the DLA/sub-DLA detected
towards QSO A (left column) and QSO B (right column) at 
$z_{\rm abs} \sim 2.66$.  The x-axis
shows a velocity scale relative to the centre of the metal line
absorption, whose redshift is given in the bottom panels. }
\end{figure}

\begin{figure}
\centerline{\rotatebox{0}{\resizebox{8cm}{!}
{\includegraphics{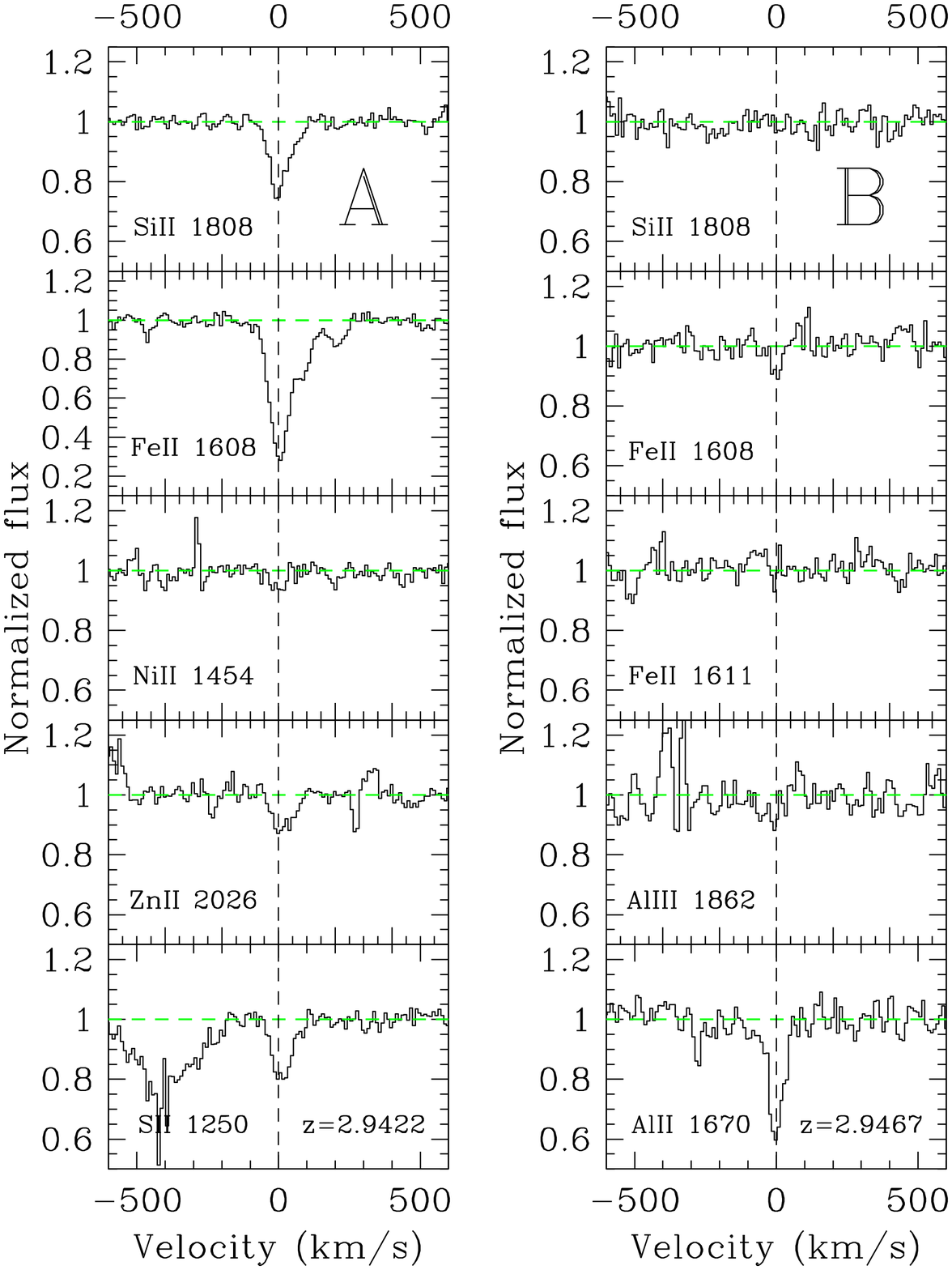}}}}
\caption{\label{met3} A selection of detected metals for the DLA/sub-DLA detected
towards QSO A (left column) and QSO B (right column) at
$z_{\rm abs} \sim 2.94$.  The x-axis
shows a velocity scale relative to the centre of the metal line
absorption, whose redshift is given in the bottom panels. }
\end{figure}

The HI column densities of the absorbers at $z_{\rm abs}$ =
2.47, 2.66 and 2.94 were determined by fitting the continuum
normalized spectra with damped profiles using the Dipso software. 
The fits to the data are shown in Figures \ref{hi_fit1}, \ref{hi_fit2}
and \ref{hi_fit3}.  For all 5 absorbers, we detect metal lines,
including Si~II, Fe~II, S~II, Zn~II and Ni~II, see Figures
\ref{met1}, \ref{met2} and \ref{met3}.  Since the resolution of ESI is
significantly larger than the Doppler widths of the metal
lines, Voigt component fitting does not yield a physically
meaningful decomposition of the line profile.
Instead, column densities are determined using the apparent optical
depth method (AODM, e.g. Savage \& Sembach 1991).    
Accurate column densities of
typically weak lines such as Zn~II can be well determined in
lower resolution spectra.  For example, the early work of Pettini
et al. (1997) determined N(Zn~II) from spectra with FWHM $\sim$
1 \AA, similar to ESI's resolution, which have subsequently been
confirmed with higher resolution echelle spectroscopy.
The AODM also allows us to assess the impact of saturation
in unresolved lines when multiple transitions from a 
given species are observed.  This is often the case for
Fe~II lines, and when saturation is suspected,
we adopt the maximum column density, usually derived
from the weakest transition.  The main disadvantage of the AODM
is that the assessment of contamination and blending is
less straightforward.

Due to their ionization potentials, it is usually assumed
that the singly ionized species represent the dominant
state of elements in DLAs.  
For absorbers with log N(HI) $\sol$ 20.0 there may be a 
non-negligible ionization
correction which undermines this assumption.  When a range
of metal lines from different ionization states are detected,
it is possible to model their relative contributions.
Often this can be done with Al~II plus Al~III or Fe~II plus
Fe~III.
However, as described below, for the two absorbers in our
spectra with log N(HI) $<$ 20.0 such an estimate of
ionization correction is not possible due to the limited
number of transitions that we detect.  In our abundance
determinations we therefore assume that N(X) = N(X~II) when
log N(HI) $>$ 20.0, but for the lower N(HI) absorbers we
quote only column densities and not abundances.

Fits to both HI and metal line
species are now described in the following sub-sections on
an absorber-by-absorber basis.  The redshifts that we
allocate to each absorber refer to zero velocity for the
metal lines, not the value determined from the Ly$\alpha$.

\subsubsection{$z_{\rm abs}$ = 2.4675 sub-DLA towards QSO B}

\begin{center}
\begin{table*}
\caption{Ionic Column Densities for $z_{\rm abs}$ = 2.4675 sub-DLA towards QSO B}
\begin{tabular}{lcccc}
\hline
Ion  & $\lambda$ & $f$-value & log N & log N$_{adopt}$ \\
\hline
H~I & 1215.6701 & 0.41640 &  & 19.85$\pm$0.15 \\
Al~II & 1670.7874 &  1.880 & 12.73$\pm$0.09 & 12.73$\pm$0.09\\
Fe~II  &  1608.4511  &  0.0580 & $<$13.6 &  13.33$\pm$0.12\\
Fe~II & 2260.7805 & 0.00244 & $<$14.2 & \\
Fe~II & 2249.8768 & 0.001821 & $<$14.4 & \\
Fe~II & 2344.2140  & 0.1140 & 13.33$\pm$0.12 & \\
Fe~II & 2374.4612 & 0.0313 & 14.10$\pm$0.09 & \\
Fe~II & 2382.7650  & 0.3200 &  13.42$\pm$0.07& \\
Si~II   &  1808.0130  &  0.002186 & $<$14.4 &  $<$14.4\\
Zn~II   &  2026.1360  &  0.4890 & 12.08$\pm$0.26 & $<$12.3\\
Zn~II   &  2062.6640  & 0.2560 & $<$12.3 &  \\
Cr~II   &  2056.2539  & 0.1050 & $<$12.6 & $<$12.6\\
\hline 
\end{tabular}\label{abund_table_z2.47}
\\ All limits are 3$\sigma$.
\end{table*}
\end{center}

Despite the low N(HI) of this absorber and the relatively poor S/N
in the blue part of QSO B's spectrum, clear damping wings 
allow us to constrain the hydrogen column density to within
0.15 dex (not including continuum fit errors).  
No corresponding DLA or sub-DLA is seen towards QSO
A, although there is a saturated Ly$\alpha$ absorber
at $z_{\rm abs} = 2.4688$ corresponding to a velocity
offset of $\Delta v$ = 95 km/s from the best fit HI redshift
of $z_{\rm abs} = 2.4677$.

Although the low N(HI) and  relatively poor S/N preclude the
detection of many metal line species, we do detect several
lines of Fe~II (although many are blended) as well as Al~II
$\lambda$ 1670 and Zn~II $\lambda$ 2026, see Table \ref{abund_table_z2.47}.  
We adopt the Fe~II
column density from the 2344 \AA\ line since the other
detected transitions of this species (Fe~II $\lambda \lambda
2374, 2382$) are blended (see Figure \ref{met1}).

One potentially serious problem with
lower resolution data is blending with either species at the
same wavelength (e.g. Zn~II $\lambda$ 2062 with Cr~II $\lambda$ 2062)
or from lines at other redshifts.  At high resolution, this is
usually easily identified during the Voigt profile fitting
process, but is not always obvious in lower dispersion data.
This issue is relevent in our determination of
a Zn~II column density, since the Zn~II $\lambda$ 2026 line
(the stronger of the two UV lines) is only separated from
Mg~I $\lambda$ 2026 by 50 km/s.  The relative strengths of Mg~I
and Zn~II will depend on a variety of factors, including intrinsic
abundance ratios, differential dust depletion and ionization structure
of the absorber.  For DLAs and sub-DLAs with
high column densities of metals, it seems that Mg~I can contribute
significantly to the total equivalent width at $\lambda \sim 2026$
(e.g. York et al. 2006; P\'eroux et al. 2006;
Herbert-Fort et al. 2006).  Although this is unlikely to
be a serious problem for this sub-DLA, whose metal
column densities are not large, we conservatively quote
an upper limit for N(Zn~II) based on the non-detection of
Zn~II $\lambda$ 2062.  Without a correction
for Mg~I and without an ionization correction (see below) the
Zn abundance determined from Zn~II $\lambda$ 2026
is high, about 0.4 $Z_{\odot}$.  It would
therefore be very interesting to improve the S/N for this absorber
to confirm (or otherwise) this relatively high metallicity.
With the limited metal column densities
we are able to reliably measure, it is not possible to
determine ionization corrections for this absorber, so
we do not attempt to convert our AODM measurements into
abundances.

It is interesting to note that although the extent of C~IV
galaxy halos are usually considered to have large sizes
(up to a few hundred kpc, Steidel et al. 1995; Adelberger
et al. 2005b), in this case, there is a C~IV absorber in QSO A
at $z_{\rm abs} = 2.4532$ compared with a redshift for C~IV in
QSO B (where the sub-DLA is detected) of $z_{\rm abs} = 2.4683$.
This corresponds to a velocity difference of $\sim 1300$ km/s.
It is not clear whether the C~IV absorption in QSO A is 
associated with the sub-DLA detected in QSO B.  The interpretation of C~IV
halo sizes clearly depends on the velocity tolerance permitted
for line matches.  We will analyse the C~IV absorbers
in this pair, and a larger sample, in a forthcoming paper.

\subsubsection{$z_{\rm abs}$ = 2.6617 DLA towards QSO A and 
$z_{\rm abs}$ = 2.6625 sub-DLA towards QSO B}

\begin{center}
\begin{table*}
\caption{Ionic Column Densities for $z_{\rm abs}$ = 2.6617 DLA towards QSO A}
\begin{tabular}{lcccc}
\hline
Ion  & $\lambda$ & $f$-value & log N & log N$_{adopt}$ \\
\hline
H~I & 1215.6701 & 0.41640 & & 20.48$\pm$0.10 \\
Al~III & 1854.7164 &  0.5390 & 12.97$\pm$0.10 & 12.97$\pm$0.10\\
Al~III & 1862.7895 &  0.2680 & 12.99$\pm$0.14 & \\
Fe~II  &  1608.4511  &  0.0580 & 14.41$\pm$0.05 & 14.36$\pm$0.10 \\
Fe~II  &  1611.2005 &  0.001360 & $<$14.5 & \\
Fe~II & 2344.2140  & 0.1140 & 14.22$\pm$0.03 & \\
Fe~II & 2374.4612 & 0.0313 & 14.30$\pm0.06$ & \\
Fe~II & 2382.7650  & 0.3200 & 14.12$\pm0.03$ & \\
Si~II   &  1808.0130  &  0.002186 & 15.05$\pm$0.11 & 15.05$\pm$0.11\\
Zn~II   &  2026.1360  &  0.4890 & 12.60$\pm0.13$ & 12.40$\pm0.20$\\
Zn~II   &  2062.6640  & 0.2560 & 12.40$\pm0.20$ & \\
Cr~II   &  2056.2539  & 0.1050 & $<$12.7 & $<$12.7\\
Ni~II & 1741.5531 & 0.04270 & 13.35$\pm$0.20 & 13.35$\pm$0.20\\
\hline 
\end{tabular}\label{abund_table_z2.66A}
\\ All limits are 3$\sigma$.
\end{table*}
\end{center}

\begin{center}
\begin{table*}
\caption{Ionic Column Densities for $z_{\rm abs}$ = 2.6625 sub-DLA towards QSO B}
\begin{tabular}{lcccc}
\hline
Ion  & $\lambda$ & $f$-value & log N & log N$_{adopt}$ \\
\hline
H~I & 1215.6701 & 0.41640 & & 20.11$\pm$0.10 \\
Fe~II & 2344.2140  & 0.1140 & 13.54$\pm$0.09 & 13.87$\pm$0.13 \\
Fe~II & 2374.4612 & 0.0313 & 13.87$\pm$0.13 & \\
Si~II   &  1808.0130  &  0.002186 &$<$14.4& $<$14.4\\
Zn~II   &  2026.1360  &  0.4890 & $<$12.1&$<$12.1\\
Cr~II   &  2056.2539  & 0.1050 &$<$12.7& $<$12.7\\
Ni~II & 1741.5531 & 0.04270 & $<$13.0& $<13.0$ \\
\hline 
\end{tabular}\label{abund_table_z2.66B}
\\ All limits are 3$\sigma$.
\end{table*}
\end{center}

For the DLA towards QSO A we detect metal lines from the following
species: Fe~II, Si~II, Zn~II, Al~III and Ni~II and determine an 
upper limit for Cr~II, see Table \ref{abund_table_z2.66A} and 
Figure \ref{met2}.  We also detect  Al~II $\lambda$ 1670,
but this transition is blended with C~IV at $z_{\rm abs} \sim 2.94$
so we do not quote a column density for it.
For the Fe~II column density, we adopt
an average of the $\lambda \lambda$ 1608 and 2374 transitions
which are the least likely to suffer from saturation.

As mentioned in the previous
subsection, Mg~I may be a significant contaminant at 2026 \AA.
The most common way to calculate
the contribution from Mg~I is to measure N(Mg~I) from the
Mg~I $\lambda$ 2852 line and predict the contribution at 2026 \AA\
according to the relative oscillator strengths.  However, at a 
redshift of $z_{\rm abs} \sim 2.66$ the Mg~I  $\lambda$ 2852 line
is shifted into the infra-red.
We therefore rely on Zn~II $\lambda$ 2062 to provide a Zn~II column
density.  Although the Zn~II $\lambda$ 2062 \AA\ line can itself suffer 
from blending with Cr~II $\lambda$ 2062, in this case the upper 
limit on the stronger Cr~II $\lambda$ 2056 rules out this possibility.
Comparing the column densities derived from the two Zn~II
lines in Table \ref{abund_table_z2.66A} shows that the contribution from Mg~I
at $\lambda$ 2026 would have led to an over-estimate of
N(Zn~II) by $\sim$ 0.2 dex.

For the sub-DLA towards QSO B, we only determine a column
density for Fe~II (see Table \ref{abund_table_z2.66B})
since Al~III $\lambda \lambda$ 1854, 1862
have very broad profiles and are slightly offset from the
central velocity, so may be blends or mis-identifications.
The usually strong Al~II $\lambda$ 1670
is also in a blended part of the spectrum.

\subsubsection{$z_{\rm abs}$ = 2.9422 sub-DLA towards QSO A and 
$z_{\rm abs}$ = 2.9467 sub-DLA towards QSO B}

\begin{center}
\begin{table*}
\caption{Ionic Column Densities for $z_{\rm abs}$ = 2.9422 sub-DLA towards QSO A}
\begin{tabular}{lcccc}
\hline
Ion  & $\lambda$ & $f$-value & log N & log N$_{adopt}$ \\
\hline
H~I & 1215.6701 & 0.41640 & & 20.28$\pm$0.05 \\
S~II & 1250.5840 &  0.005453 & 15.01$\pm$0.10 & 15.01$\pm$0.10\\
S~II & 1253.8110 &  0.01088 & 15.01$\pm$0.08 & \\
Al~II & 1670.7874 &  1.880 &13.84$\pm$0.06 & $\ge$13.84 \\
Al~III & 1854.7164 &  0.5390 & 13.73$\pm$0.04 & $\ge$13.73\\
Al~III & 1862.7895 &  0.2680 &13.73$\pm$0.07 & \\
Fe~II  &  1608.4511  &  0.0580 &14.69$\pm$0.04 & 14.69$\pm$0.04\\
Fe~II  &  1611.2005 &  0.001360 & 14.66$\pm$0.21 & \\
Si~II   &  1808.0130  &  0.002186 & 15.34$\pm$0.08 & 15.34$\pm$0.08\\
Zn~II   &  2026.1360  &  0.4890 & 12.62$\pm$0.12 & 12.40$\pm0.33$\\
Ni~II & 1370.1310 &  0.07690 &13.71$\pm$0.23 & 13.66$\pm$0.28 \\
Ni~II & 1454.842 & 0.0323 & 13.60$\pm$0.21 & \\
\hline 
\end{tabular}\label{abund_table_z2.94A}
\\ All limits are 3$\sigma$.
\end{table*}
\end{center}

\begin{center}
\begin{table*}
\caption{Ionic Column Densities for $z_{\rm abs}$ = 2.9467 sub-DLA towards QSO B}
\begin{tabular}{lcccc}
\hline
Ion  & $\lambda$ & $f$-value & log N & log N$_{adopt}$ \\
\hline
H~I & 1215.6701 & 0.41640 & & 19.70$\pm$0.10 \\
S~II & 1253.8110 &  0.01088 & $<$14.0 & $<$14.0\\
Al~II & 1670.7874 &  1.880 &12.57$\pm$0.10 &12.57$\pm$0.10\\ 
Al~III & 1862.7895 &  0.2680 &12.51$\pm$0.22 & $<12.6$\\
Fe~II  &  1608.4511  &  0.0580 & 13.29$\pm$0.25 & 13.29$\pm$0.25 \\
Si~II   &  1808.0130  &  0.002186 & $<$14.6 & $<$14.6\\
Zn~II   &  2026.1360  &  0.4890 &$<$12.1 & $<$12.1\\
Ni~II & 1370.1310 &  0.07690 &$<$13.4 & $<$13.4\\
\hline 
\end{tabular}\label{abund_table_z2.94B}
\\ All limits are 3$\sigma$.
\end{table*}
\end{center}

The best fit N(HI) to the $z_{\rm abs} \sim$ 2.94 absorber in
QSO A yields a column density just below the DLA threshold,
although within the error bars it may still be a `classical'
DLA.  The absorber towards QSO B appears to be a blend,
but the red damping wing allows a reasonable HI fit that
classifies this absorber as a sub-DLA.

For the  $z_{\rm abs}$ = 2.9422 sub-DLA towards QSO A we detect
a range of metal species: Si~II, Fe~II, Zn~II, S~II, Ni~II,
Al~II and Al~III, see Table \ref{abund_table_z2.94A} and Figure \ref{met3}.  
N~V $\lambda$ 1242 may be present, but is weak
(peak optical depth $\tau \sim$ 0.1) and blended with another
feature.  We do not quote a detection or limit for Cr~II
since all the lines are in a contaminated part of the spectrum.
Al~II $\lambda$ 1670 is clearly saturated, so we only quote a lower
limit for N(Al~II).  For Al~III $\lambda
\lambda$ 1854, 1862 the two lines give consistent column densities,
despite their high equivalent widths.  However, to be cautious, we quote
their column densities as lower limits.
The Fe~II $\lambda$ 1608 line towards QSO A is quite strong
and may be slightly
saturated.  However, the  much weaker Fe~II $\lambda$ 1611 is marginally
detected and gives a consistent column density.  

We have the same problem for Zn~II as
for the DLA at $z_{\rm abs} \sim 2.66$ discussed in
the previous subsection, i.e. the potential contribution
from Mg~I to the Zn~II $\lambda$ 2026 line.  However,
the situation is even more complicated in this case because
not only is Mg~I $\lambda$ 2852 not covered, but also the
Zn~II $\lambda$ 2062 line is in a highly contaminated part of the
spectrum, so can not be used and we have no estimate of the
Cr~II contribution.  
We therefore estimate an upper limit to the contribution
of Mg~I by considering values measured in other metal-rich
absorbers.  From P\'eroux et al. (2006) and Herbert-Fort et al. (2006),
we assume a rest-frame upper limit contribution of 50 m\AA\ 
from Mg~I.  This means that N(Zn~II) may require a downward
revision of up to 0.4 dex.  We therefore quote a column density
of Zn~II with error bars that account for this range of possibilities,
see Table \ref{abund_table_z2.94A}.  Regardless of our uncertainties
in N(Zn), this absorber is clearly metal rich, since
the column densities of Si and S also yield abundances $\sim$ 1/3
$Z_{\odot}$.  

We also search for H$_2$ in the sub-DLA towards QSO A.
At $z_{\rm abs}$ = 2.9422 we have coverage of Lyman 
$J=0, 1$ rotational bands for $6-0$ to $0-0$ vibrational
transitions.  These first two $J$ states usually dominate
the molecular column density in DLAs (e.g. Ledoux et al. 2003).
Assuming a redshift matched to the central
position of the metal lines, i.e. $z_{\rm abs} = 2.9422$,
we combine the limits on $J = 0$ and $J = 1$ to determine a limit 
of log N(H$_2$) $<$ 14.5 in this line of sight.

For the sub-DLA towards QSO B, we only determine column
densities for Fe~II and Al~II and upper limits for the other
species listed in Table \ref{abund_table_z2.94B}.  Al~III
$\lambda$ 1854 is in a region of moderate contamination
in the B spectrum, so that the column of a weak line is
unreliable. There is
a marginally significant (2.6$\sigma$) feature close to the
expected position of Al~III $\lambda$ 1862, but since it is
below the 3$\sigma$ level, we quote it as a limit.

\begin{center}
\begin{table*}
\caption{Abundances for Absorbers with N(HI) $>$ 20.1}
\begin{tabular}{lcccccccc}
\hline
QSO & $z_{\rm abs}$ & Log N(HI) & [Fe/H] & [Si/H] & [Zn/H] & [Cr/H] & [S/H] & [Ni/H] \\ 
\hline
A & 2.6617 & 20.48$\pm$0.10 & $-1.59\pm0.14$ & $-0.97\pm0.15$ & $-0.71\pm0.22$ & $<-1.43$ & ... & $-1.35\pm0.22$ \\
B & 2.6625 & 20.11$\pm$0.10 & $-1.71\pm0.16$ & $<-1.25$ & $<-0.51$ & $<-1.06$& ... & $<-1.33$\\
A & 2.9422 & 20.28$\pm$0.05 & $-1.06\pm0.06$ & $-0.48\pm0.09$ & $-0.51\pm0.33$ & ... & $-0.46\pm0.11$ & $-0.84\pm0.28$\\
\hline 
\end{tabular}\label{abund_table}
\\ All limits are 3$\sigma$.
\end{table*}
\end{center}

\section{Discussion}

\subsection{Abundances}

The final abundances for the three absorbers with log N(HI) $>$ 20
are given in Table \ref{abund_table}.
The 2 absorbers in QSO A both have very high metallicities, based
on the abundance of Zn: 1/3 and 1/5 $Z_{\odot}$.
Such high metallicities are very rare even in low redshift DLAs
(e.g. Meiring et al. 2006), leading to the widespread conclusion
that the DLA cross-section is dominated by metal-poor gas.  It has
been suggested that this may be due to a dust-induced bias against
metal-rich galaxies, although there is currently no observational
evidence to support this claim (e.g. Ellison et al. 2001, 2004, 2005;
Akerman et al. 2005; Jorgenson et al. 2006).  Herbert-Fort et al. (2006)
have shown that selecting absorbers from the SDSS on the basis of
strong metal lines can readily identify DLAs with metallicities
in excess of $0.1 Z_{\odot}$, indicating that metal-rich systems
do exist at high redshift, but are simply rare.  The two DLAs studied here have
metallicities consistent with the metal-strong population of
Herbert-Fort et al.  Moreover, the abundances that we determine
are comparable to those measured in cB58 (2/5 $Z_{\odot}$ at $z=2.7$, 
Pettini et al. 2002) and other LBGs at $2 < z < 3$
(Teplitz et al. 2000; Shapley et al. 2004).  

We also determine relatively high depletion factors for the DLAs based
on [Zn/Fe] = +0.88, +0.55 in the DLAs at $z_{\rm abs} = 2.66$ and
2.94 respectively.  These values are usually considered as measures
of the dust:metals ratio, since these two elements trace each other
well in most Galactic stars (although see the caveats in Nissen et al.
2004),
yet Fe is highly refractory whereas Zn is not (Savage \& Sembach 1996).
The relative abundances measured here imply that $\sim$ 15 -- 30 \%
of the metals in these two DLAs are in the gas phase and follows
the broad trend of increased depletion with increasing metallicity
(e.g. Meiring et al. 2006 for the most recent compilation).
Although quite extreme by DLA standards, these depletion factors still barely
overlap with measurements in the Galactic disk and Magellanic
Clouds (e.g. Roth \& Blades 1997).

Although this sightline intersects gass associated with the diffuse
ISM, given its high metallicity (1/3 $Z_{\odot}$) and high depletion, 
it is perhaps
surprising that we do not detect H$_2$ down to a molecular fraction
of log $f$(H$_2$) = 2N(H$_2$) / [2N(H$_2$) + N(HI)] $< -5.5$
for the $z_{\rm abs} = 2.94$ DLA\footnote{Although it
is possible that we have under-estimated the
contribution of Mg~I to the Zn~II $\lambda$ 2026 line and the
true metallicity may be less than we have deduced.  However, the
abundance based on Si or S is also 1/3$Z_{\odot}$.}.  Ledoux et al. (2003) and
Petitjean et al. (2006) have suggested that both metallicity
and depletion may be important factors for a galaxy to be able to
form (and maintain) a significant column density of molecules.
As a comparison with the $z_{\rm abs} = 2.94$ DLA studied here,
we can consider molecular fractions in LMC sightlines, where the metallicity
is similar to this DLA.  Tumlinson et al. (2002) showed
that H$_2$ is detected in $\sim$ 50\% of LMC sightlines, so we
may naively expect a similar detection rate in other galaxies
with similar metallicities.  Although the statistics for
metal-rich DLAs are still poor, a 50\% detection rate does
broadly fit the high redshift data for DLAs with metallicities
above 1/5 $Z_{\odot}$ (Petitjean et al. 2006).  However,
metallicity is clearly not the only factor, since the SMC,
whose metallicity is a factor three lower than the LMC,
has detections for 90\% of the sightlines studied by Tumlinson
et al. (2002).  This has been proposed to be due to higher
star formation rates in the LMC which preferentially photo-dissociate
molecules.  

Finally, we note that the $z_{\rm abs} = 2.94$ DLA has a relative 
abundance of S and Zn that is close to the solar value, [S/Zn] = $+0.05$.  
In Galactic stars, we observe enhanced $\alpha$/Fe
ratios at metallicities below [Fe/H] $< -1.0$, due to the
time delay between SNII and SNIa enrichment.   In DLAs,
the two undepleted elements S (an $\alpha$ element) and Zn
(which traces the Fe-peak in Galactic halo stars
at [Fe/H] $> -2.5$) are usually preferred tracers of the relative
contributions of SNIa and SNII (e.g. Nissen et al. 2004).  The universally
low (typically sub-solar) values of [S/Zn] in DLAs indicates that
their chemical enrichment history is very different to that
of the Milky Way.   New measurements from local dwarf galaxies
indicate that low $\alpha$/Fe ratios may actually be common in the
Local Group, e.g. Shetrone et al. (2003).  At a metallicity of 1/3$Z_{\odot}$
the value of [S/Zn] measured here is actually intermediate
between typical Galactic values (e.g. Gratton et al. 2003)
and those found in the more metal rich Local Group dwarfs such
as Fornax, the LMC and Sagittarius (Pompeia, Hill \& Spite 2005;
Bonifacio et al. 2004).

\subsection{Are DLA galaxies over 100 \hkpc\ in size?}\label{sim_sec}

At low redshift, we have a number of measurements of DLA absorber
size.  At $z = 0$,
Zwaan et al. (2005) showed that a 21cm survey
of selected UGC galaxies finds typical dimensions for DLA-equivalent 
column densities ranging from $\sim$ 5 -- 50 kpc, depending
on galaxy mass and orientation.  At similarly low redshifts
($cz < 4000$ km/s) 
Bowen, Pettini \& Blades (2002) found that if a line of sight passes
within 200 kpc of a nearby galaxy, the covering factor for
HI gas with log N(HI) $>$ 13 is effectively 100\%.  However,
the cross-section of gas that is optically thick at the
Lyman limit will be considerably smaller than for absorbers
with  13 $<$ log N(HI) $<$ 16.  At these lower column densities,
the HI absorption can presumably be associated not only with extended,
diffuse halos, but also with non-galactic
structures such as filaments and intra-group media.  
At intermediate redshifts, Chen \& Lanzetta (2003) determined
a characteristic size for DLAs of $R_{\star} \sim$ 25 kpc,
and also found that a significant number of DLA fields have multiple
galaxies at the absorber redshift.    In contrast,
constraints on galaxy/absorber size on kpc scales at $z > 2$ are very few.  
They are limited to 2 pairs of QSO sightlines which probe
transverse scales of 5--10 kpc (Lopez et al. 2005; Smette et al. 1995), 
a measurement of
spatially offset Ly$\alpha$ fluorescence (Adelberger et al. 2006),
a handful of extended radio sources (Foltz et al. 1988;
Briggs et al. 1989) and extended [OIII] emission
from 3 DLAs at $2 < z < 3$ (M\o ller, Fynbo \& Fall 2004;
Weatherley et al. 2005).  Although these
observations generally imply gaseous extents of at most a few
tens of kpc, Labb\'e et al. (2003) have used deep imaging to
detect a population of galaxies at $1.5 < z <3$ whose disks have 
a comparable size to the Milky Way and Prochaska \& Wolfe (1997)
have also interpreted DLAs as large disks.  The critical
question is therefore: is it likely that the common
absorption seen in SDSS 1116+4118 AB is due to galaxies 
with 100 kpc gas halos at $z \sim 3$? 

Some first clues as to whether the coincident absorption
is due to a common structure can be gleaned from the relative velocities
across the line of sight.  Based on the metal lines, the velocity
separations between the DLA/sub-DLA in QSO A/B are 65 km/s ($z_{\rm abs} \sim
2.66$) and 340 km/s ($z_{\rm abs} \sim 2.94$).    Although only
a handful of rotation curves and velocity fields have been studied
at high redshift (e.g. Erb et al. 2003, 2004), a significant
number have velocity dispersions in excess of 200 km/s and
rotation velocities $> 150$ km/s (i.e. a total velocity difference
$>$ 300 km/s).  Therefore, although the shear between QSO A/B is large
for the $z_{\rm abs} \sim 2.94$ absorber, it is still feasible that
it is within the same galaxy.

We can also appeal to the metallicities for clues to the absorbers'
structure.  At both $z \sim 2.66$ and $z \sim$ 2.94, at least
one of the absorbers has a high metallicity.  On the one hand,
the existence of a mass-metallicity relation at all redshifts
studied (Tremonti et al. 2004; Savaglio et al. 2006; Erb et al.
2006) indicates that these galaxies may have stellar masses
at least $\sim$ $10^9 - 10^{9.5}$ M$_{\odot}$, possibly lending support
to the idea of a large physical extent.  On the other hand, such
high abundances seen in absorption, consistent with metallicities
inferred from the emission line measurements in LBGs, indicates 
that these sightlines may intersect the central part of their
galaxies.  If true, it 
would imply even larger sizes for the absorbing
galaxy than the simple interpretation of common coincidence, 
i.e. $r \sog$ 100 \hkpc, rather than $d \sog$ 100 \hkpc.  However,
neither the information on metallicity or velocity provide
very satisfactory conclusions as to the nature of the coincident
absorption. We therefore turn to galaxy simulations to help us interpret
our observations.

\begin{figure*}
\centerline{\rotatebox{0}{\resizebox{16cm}{!}
{\includegraphics{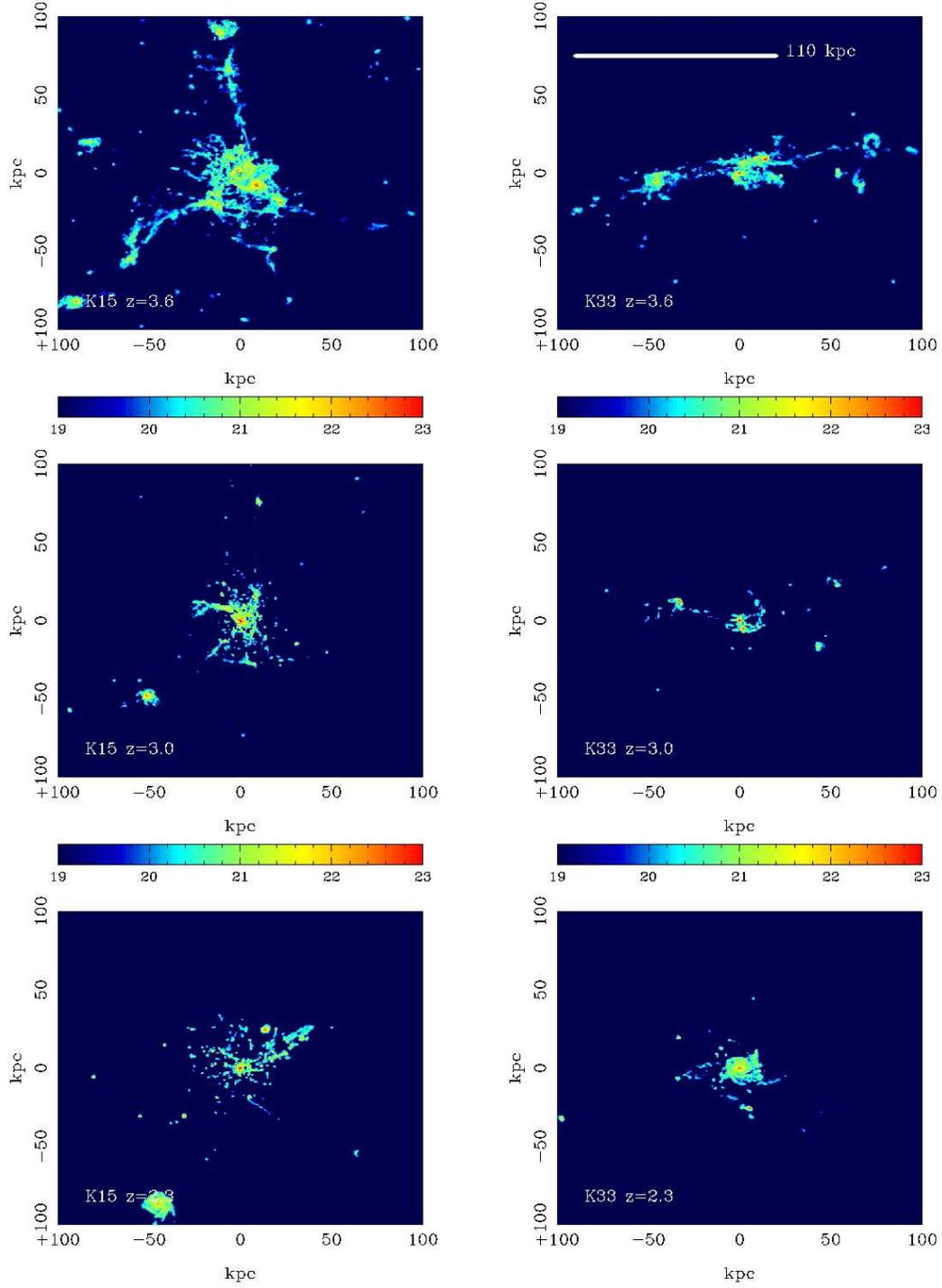}}}}
\caption{\label{sim_box_xy_200} Projected N(HI) map for the central
200 kpc for our two simulated galaxies, K15 (left) and K33(right) 
at $z=3.6, 3.0$ and $2.3$ (top through bottom panels).
This projection is through the XY (face-on) plane.  The bar in the top
right panel shows the typical transverse separation at the absorber
radshift for the QSO pair studied here.  The white bar in the top
right panel shows 110 \hkpc, the approximate physical separation
of the 2 lines of sight at the absorbers' redshift in SDSS1116+4118 AB.}
\end{figure*}

\begin{figure*}
\centerline{\rotatebox{0}{\resizebox{16cm}{!}
{\includegraphics{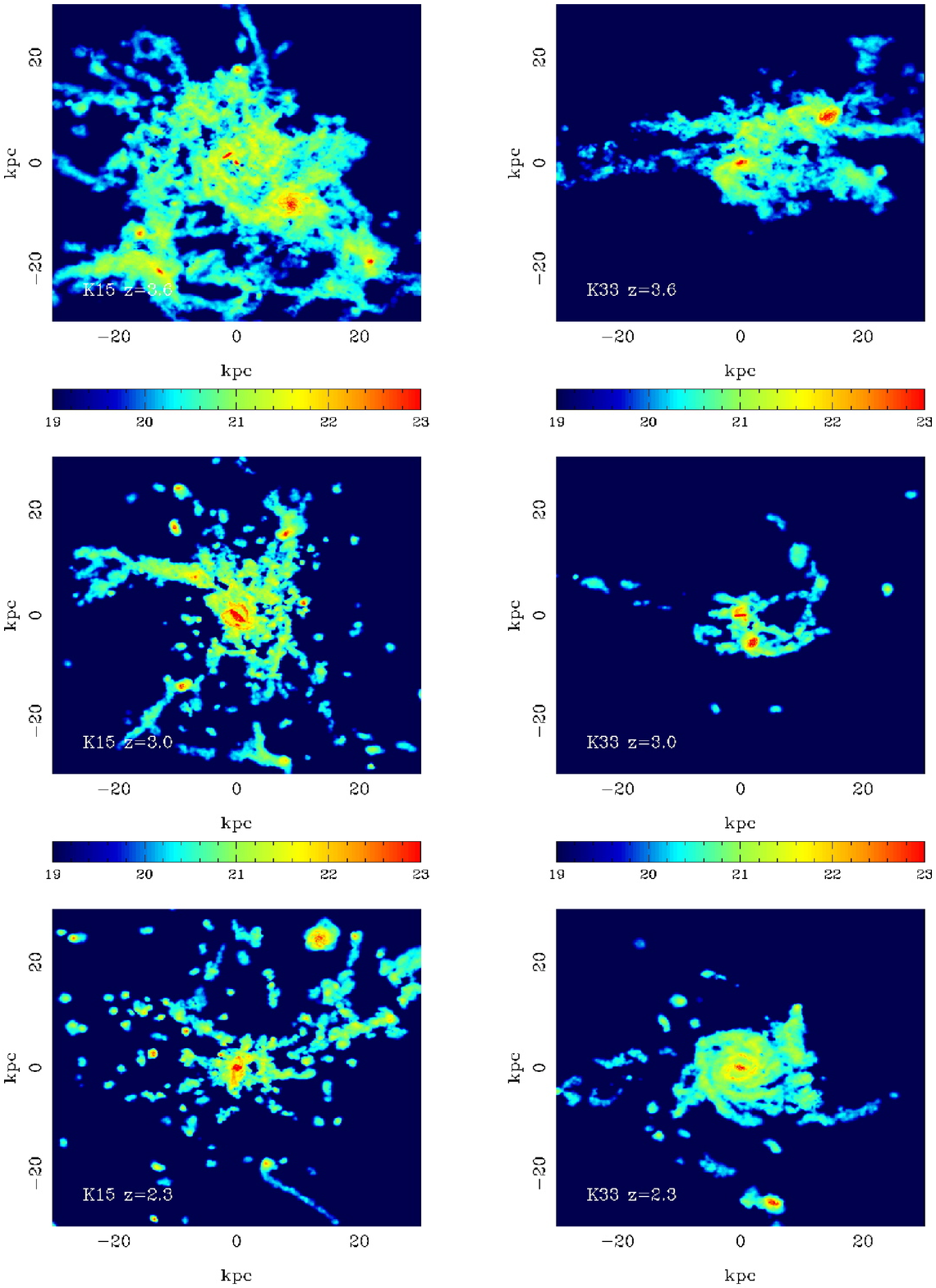}}}}
\caption{\label{sim_box_xy_60} Projected N(HI) map for the central
60 kpc for our two simulated galaxies, K15 (left) and K33(right) 
at $z=3.6, 3.0$ and $2.3$ (top through bottom panels).
This projection is through the XY (face-on) plane.  }
\end{figure*}

\begin{figure*}
\centerline{\rotatebox{0}{\resizebox{16cm}{!}
{\includegraphics{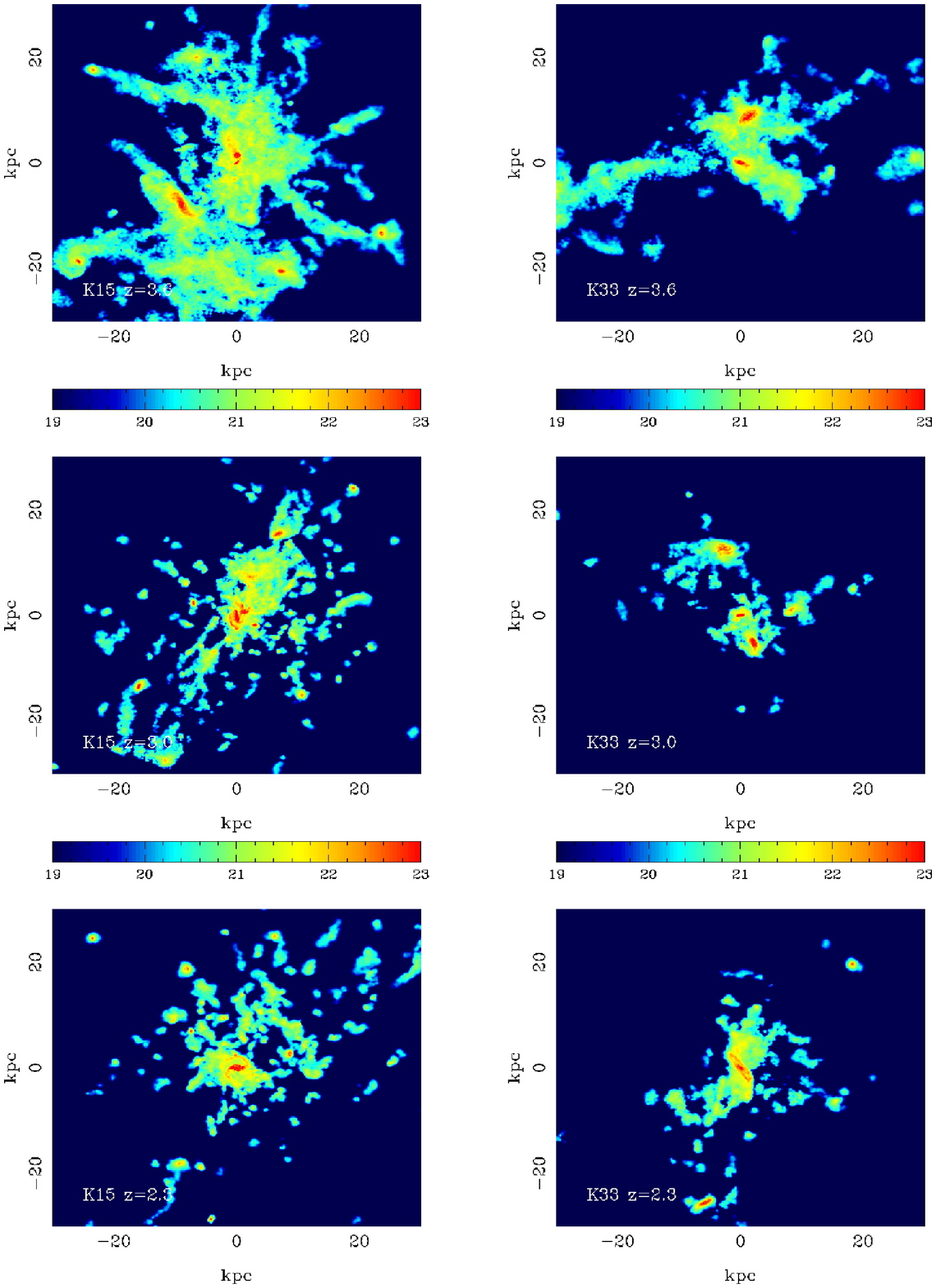}}}}
\caption{\label{sim_box_xz_60} Projected N(HI) map for the central
60 kpc for our two simulated galaxies, K15 (left) and K33(right) 
at $z=3.6, 3.0$ and $2.3$ (top through bottom panels).
This projection is through the XZ (edge-on) plane.  }
\end{figure*}

\begin{figure*}
\centerline{\rotatebox{270}{\resizebox{16cm}{!}
{\includegraphics{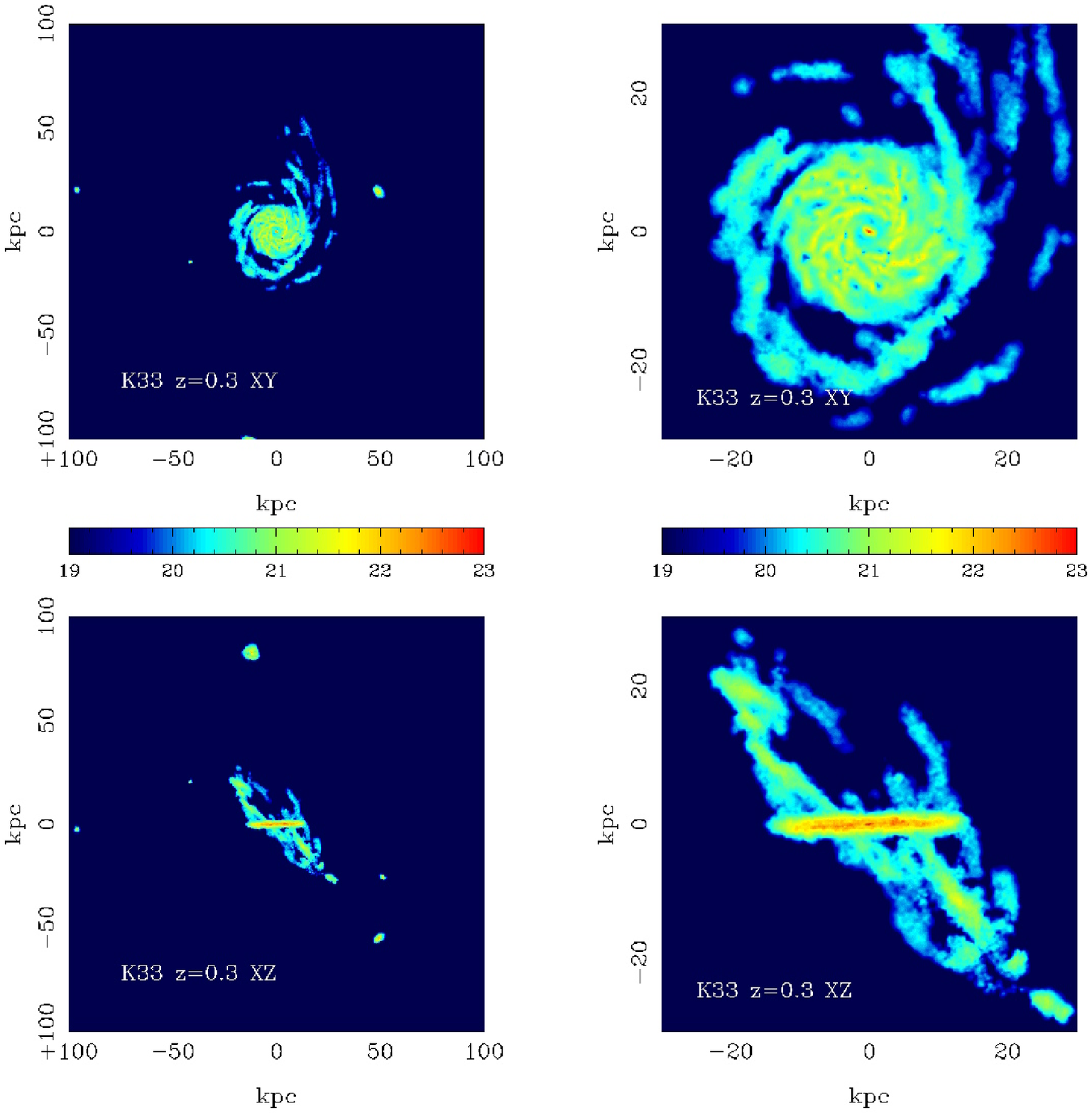}}}}
\caption{\label{sim_box_z0.3} Projected N(HI) map for the lowest
redshift slice currently simulated for galaxy K33.  Two scales
and two orientations are shown, as for Figures \ref{sim_box_xy_200}
to \ref{sim_box_xz_60}.  The specific angular momentum, $j$, of the galaxy is
660 \kms\ kpc, near the median of the observed range of $j \sim 350-1400$
\kms\ kpc typical for a disk galaxy of $V_c$=180 \kms\ at $z\sim0$ (e.g.,
Sommer-Larsen et al. 2003).  }
\end{figure*}

\subsubsection{Cosmological simulations of galaxy formation}

The code used for the simulations is a significantly improved version of
the TreeSPH code, which has been used previously for galaxy formation 
simulations (Sommer-Larsen, G\"otz \& Portinari 2003).
The main improvements over the previous version are:
(1) The `conservative' entropy equation solving scheme (Springel \& 
Hernquist 2002). 
(2) Non-instantaneous gas recycling and chemical evolution, tracing
10 elements (Lia et~al.\ 2002a,b); the algorithm includes 
supernov\ae\ of type II and type Ia, and mass loss from stars of all masses.
(3) Atomic radiative cooling depending both on the metal abundance
of the gas and on the meta--galactic UV field, modeled after Haardt
\& Madau (1996) is invoked, as well as simplified treatment
of radiative transfer, switching off the UV field where the gas
becomes optically thick to Lyman limit photons on scales of $\sim$ 1~kpc.

In this paper, very high resolution simulations of two galaxies, known to
become large disk galaxies at $z$=0, are analysed in relation to DLA and 
sub-DLA sight-line properties. We take the two galaxies simulated
by Razoumov \& Sommer-Larsen (2006), K15 and K33, which have, at $z$=0,
characteristic circular speeds of $V_c$=245 and 180 km/s, respectively.
The galaxies bracket typical disk galaxy formation histories: 
the formation of the larger
K15 disk is merger induced (e.g., also Robertson et al. 2004), with the
disk growing strongly between $z$=1 and 0, whereas K33 already
starts developing a disk by $z \sim 2.5$, which subsequently 
grows gradually to the present
epoch. The $z$=0 virial masses of the two galaxies are 
$M_{\rm{vir}}=8.9$x10$^{11}$ and 3.7x10$^{11} \Msun$, for K15 and
K33 respectively.

The two galaxies were simulated in the standard $\Lambda$CDM cosmology
using the `zoom-in' technique (e.g., Sommer-Larsen 2006) 
to study the formation and evolution of 
individual galaxies in full cosmological context. Total particle numbers
used were 2.2 and 1.2 million, for K15 and K33 respectively. 
Particle masses and 
gravitational (spline) softening lengths of  
$m_{\rm{gas}}$=$m_*$=8.45x10$^4$ and $m_{\rm{DM}}$=4.83x10$^5$ 
$h_{70}^{-1}$M$_{\odot}$, and $\epsilon_{\rm{gas}}$=$\epsilon_*$=176 and 
$\epsilon_{\rm{DM}}$=316 $h_{70}^{-1}$pc, respectively, were adopted.
The
gravity softening lengths were fixed in physical coordinates from $z$=6
to $z$=0, and in co-moving coordinates at earlier times. The minimum
SPH smoothing length in the simulation was about $12 h_{70}^{-1}$ pc.
A Kroupa IMF (Kroupa 1998) was used in the simulations, and early rapid and 
self-propagating star-formation (sometimes dubbed `positive feedback')
was invoked (Sommer-Larsen et al. 2003). 

Although K15 and K33 galaxies have been chosen from
our simulation to represent two different disk galaxy evolutionary
paths (and end products) there are some generic features
to both galaxies.  For example,
they both tend to be larger and have a more extended
gas distribution at higher redshift.  At low redshift,
the gas has cooled and settled into a more organised and
centrally concentrated structure.  However, proto-disks
can form even at quite high redshifts (see the top left panel
of Figures \ref{sim_box_xy_200} and \ref{sim_box_xy_60}). 
The two different scales for these figures highlight that
the majority of DLA gas is distributed on scales of tens,
rather than hundreds, of
kpc.  Moreover, the region of the galaxy which
is actively star formating is very small ($\sim$ kpc scale)
compared with the distribution of high column density gas.
This is qualitatively consistent with the observation of
low abundances in DLAs (e.g. Pettini et al. 1999) compared
with emission line abundances in absorption selected galaxies
(e.g. Ellison, Kewley \& Mallen-Ornelas 2005; Chen, Kennicutt
\& Rauch 2005).  The spatial separation of the bulk of
DLA-inducing gas from the regions of high star formation has
also been inferred by the lack of low surface brightness galaxies in the
Hubble ultra deep field (Wolfe \& Chen 2006). 

We analyse the model galaxies by passing sightlines through the
simulation box and integrating the HI volume density along the
line of sight.  
In Figures \ref{sim_box_xy_200} to \ref{sim_box_xz_60}
we show two projections of the two simulated
galaxies where each box has been projected into a 600 x 600
grid. The colour contours show the projected HI
gas column density in units of atoms cm$^{-2}$ and are therefore
analogous to the observed N(HI) columns measured in QSO sightlines.
The N(HI) at each point along the line of sight
is evaluated by weighting all of the particles within the local
smoothing length $h$ (where $2h$ is the distance within which 50
particles are located) according to the spline smoothing kernel 
(Monaghan \& Lattanzio 1985).
For this work, we use two different renderings of each galaxy
in order to alleviate orientation effects.  These two
projections correspond to projecting lines of sight through the
XY and XZ faces of the same simulation cube.  The XY plane corresponds
roughly to a face-on configuration of the central star-forming disk
whereas the XZ is approximately edge-on. However, as can be
seen from a visual inspection of Figures  
\ref{sim_box_xy_200} to \ref{sim_box_xz_60}, the distribution
of HI gas is not significantly different even in these two
`extreme' orientations; this is checked quantitatively below.
Although we do not use simulations at lower redshifts, for
demonstration pruposes we include here images of the K33 at $z=0.3$,
the lowest redshift to which the simulation has currently progressed.
Figure \ref{sim_box_z0.3} shows the galaxy on two spatial
scales and in both the face-on and edge-on orientations.  These images
show the stability and growth of the disk since $z=3.6$.
The K15 simulation has not yet progressed beyond $z=2.0$, so
we can not show a lower redshift rendition.

\begin{figure*}
\centerline{\rotatebox{270}{\resizebox{10cm}{!}
{\includegraphics{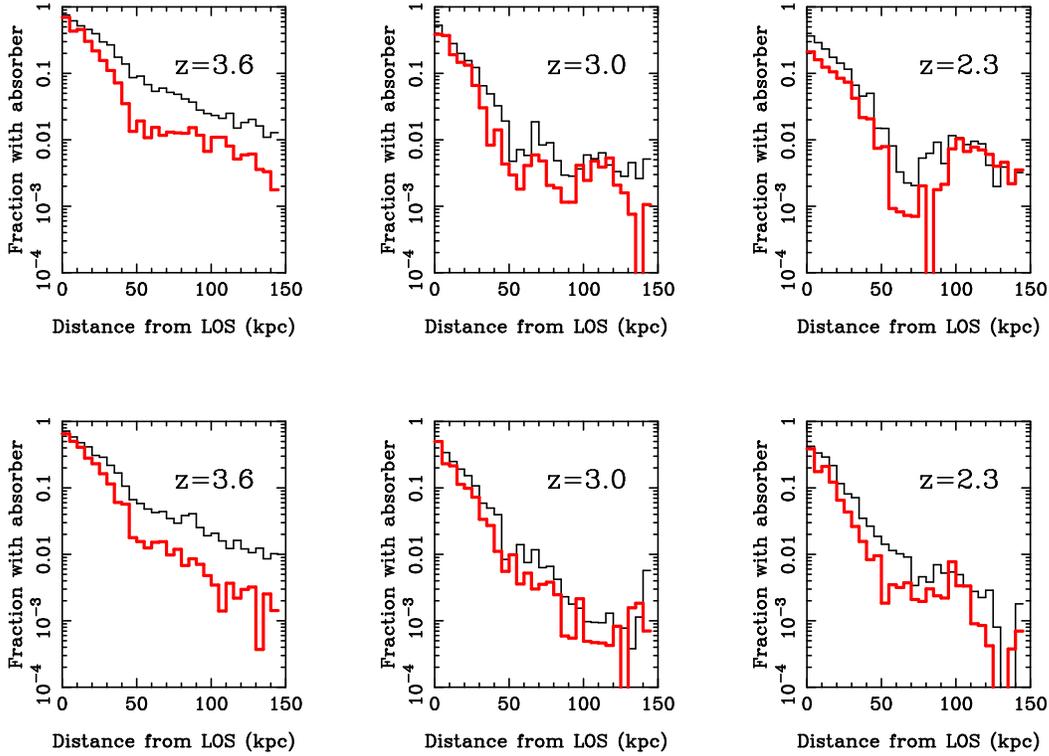}}}}
\caption{\label{double_K15} The fraction of second sightlines in a pair, 
as a function of separation, that will exhibit a DLA (red) or 
sub-DLA (black) given the presence of a DLA or sub-DLA in the first.
This probability function for finding common absorption is
calcuated for 3 different redshifts (see panel labels) for
the more massive galaxy in our simulation, K15.  The top three
panels are for the XY (face-on) orientation, and the bottom
three panels are for the XZ (edge-on) orientation.  The pair
probabilities for the two orientations are very similar, as
may be expected from a visual inspection of Figures 
\ref{sim_box_xy_200} to \ref{sim_box_xz_60}.
}
\end{figure*}

\begin{figure*}
\centerline{\rotatebox{270}{\resizebox{10cm}{!}
{\includegraphics{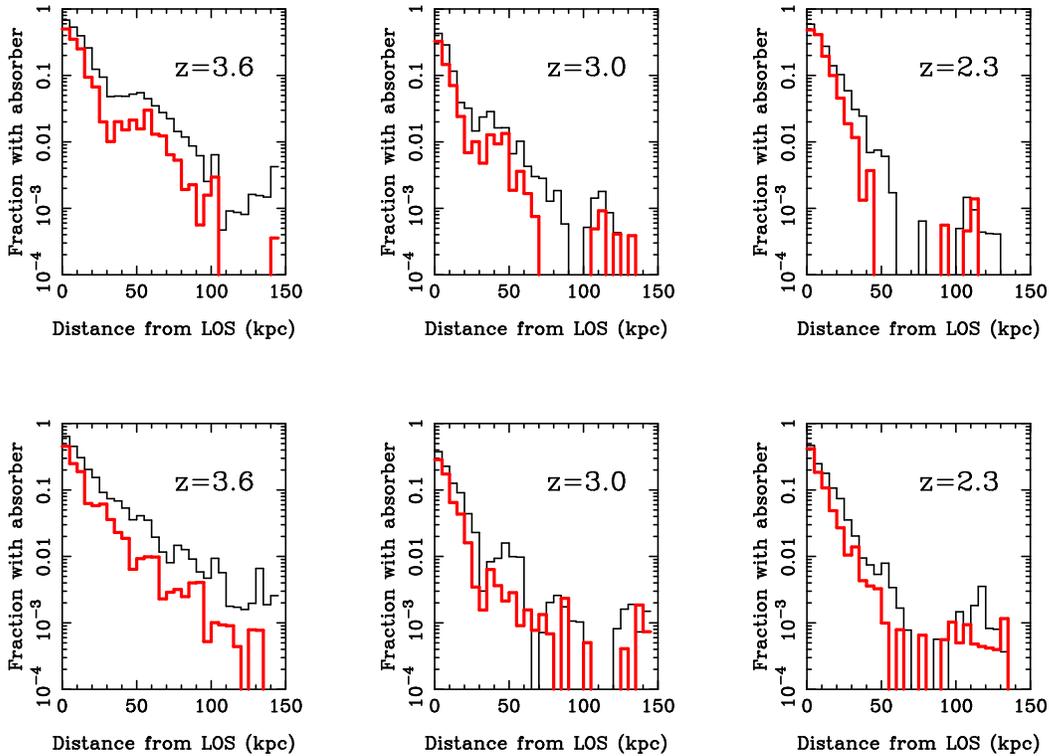}}}}
\caption{\label{double_K33} As for Figure \ref{double_K15} but for
the less massive galaxy in our simulation, K33.
}
\end{figure*}

In order to calculate the probability of finding a DLA or sub-DLA
in each of a pair of sightlines, we construct a sample of 200,000
pairs from each of the simulated galaxies at $z = 2.3, 3.0$ and 3.6.  
This suite includes
100,000 sightlines from each of the face-on and edge-on renditions.  
In every pair, the first sightline
is drawn at random from the sub-set known to intersect sub-DLA 
(log N(HI) $\ge$ 19.5) or DLA (log N(HI) $\ge$ 20.3) gas.
The second sightline is drawn at random from the full simulation.
In Figure \ref{double_K15} and \ref{double_K33} we 
show the fraction of pairs with a sub-DLA or DLA in the second line of
sight, given the presence of a sub-DLA or DLA in the first for galaxies
K15 and K33\footnote{
In this simulation we classify sub-DLA as absorbers with log N(HI) $\ge$
19.5 gas, i.e. grouping together classical DLAs and sub-DLAs into
the same category.}.  
For example, for the larger K15 galaxy at $z=3.0$
about 6\% of sightlines with a DLA in the first
line of sight also have a DLA in the second for separations
20 -- 30 kpc.  For sub-DLAs, this fraction increases to $\sim 10$\%.
At separations from 40 to 150 kpc, the likelihood of finding
coincident DLA absorption decreases from $\sim$ 1\%
to 0.1\% in K15.  In the smaller K33 galaxy, these probabilities
are about a factor of 5 smaller at $z=3.0$ and $z=3.6$.

In our observations of SDSS 1116+4118 AB we have one case of a sub-DLA
with no counterpart in the second line of sight ($z_{\rm abs} = 2.47$), 
one case of a DLA with sub-DLA counterpart ($z_{\rm abs} = 2.66$)
and one case of a pair of sub-DLAs ($z_{\rm abs} = 2.94$) all
with separations $\sim$ 110 \hkpc.  From Figures \ref{double_K15}
and  \ref{double_K33}
it can be seen that the probability of these two `double hits'
is $<10^{-5}$.  The present 
simulations still fail to reproduce the observed angular momenta
of actual disk galaxies by about a factor of 1.5, and hence the
real (proto-)disk galaxies may be up to ~50\% more extended than 
the present models.  However, the probability of a double intersection
is still $<$ 1\% at 110/1.5 $\sim$ 75 \hkpc.
Another potential uncertainty in the models is our simplified treatment of
radiative transfer, which governs the transition from neutral to
ionized gas and can therefore affect the effective area of projected sub-DLA 
absorbers.  Experiments with our radiative transfer treatment
indicate that this effect is likely to affect sub-DLA numbers by,
at most, a factor of two.  Simulations that contain a more complete
radiative transfer calculation are currently underway and will
be presented in a future paper (see also Razoumov et al. 2006).  
Even with the uncertainties
of angular momentum and the transition between neutral and ionized
gas,  the conclusion
that a single large galaxy, such as those tested here, 
is unlikely to cause coincident
absorption in SDSS 1116+4116 is unchanged, although tests with
a larger suite of galaxies would be desirable.
As the two simulation volumes also include satellites and 
merging  components around the (proto-)galaxies (e.g.
bottom left panel of Figure \ref{sim_box_xy_200}), the low  
probability of coincident pairs found above
also argues against a high covering fraction of
satellite systems or an actively merging galaxy
(such as that observed by Miley et al. 2006)
as the explanation for the common absorption across 
the line of sight, at least based on these two simulated galaxies.
Again, a wider suite of simulations is needed to confirm this
for a more extensive parameter space.

\subsubsection{DLA clustering analysis}

Having rejected a large single system, and a galaxy plus
satellite structure as the reason for coincident absorption
towards SDSS 1116+4118 AB, we now consider the possibility
 that we are
observing a structure that contains multiple galaxies.
Such a possibility is perhaps not completely unexpected
given that galaxy groups represent the most common 
environment at least at low redshift (e.g Tully 1987).
Galaxy clustering at $z \sim 3$ is becoming increasingly
well-documented, such as the extensive literature on LBGs
e.g. Giavalisco et al. (1998), Steidel et al. (1998), Porciani 
\& Giavalisco (2002), Adelberger et al. (2005a,c), Steidel
et al. (2005).
Although these investigations have selected galaxies based
on their emission properties, Cooke et al. (2006) have
shown that DLAs have similar spatial distributions as the
LBGs.  DLAs may therefore also be highly clustered at $z \sim 3$.
This posit is supported by recent observations of
Zibetti et al. (2007) who stack images of the QSO fields
of SDSS-selected Mg~II
absorbers.  They find extended starlight out to $\sim$ 200 kpc,
indicating that there can be galaxies at the redshift of
the absorber out to these impact parameters, consistent with
a group environment.  Chen \& Lanzetta (2003) have also found
that the fields of DLAs at $z<1$ often have multiple galaxies
at the absorber redshift.   We can estimate the likelihood
of a multiple galaxy coincidence by considering the clustering
scale of DLA galaxies and their column density distribution.

\begin{figure}
\centerline{\rotatebox{0}{\resizebox{10cm}{!}
{\includegraphics{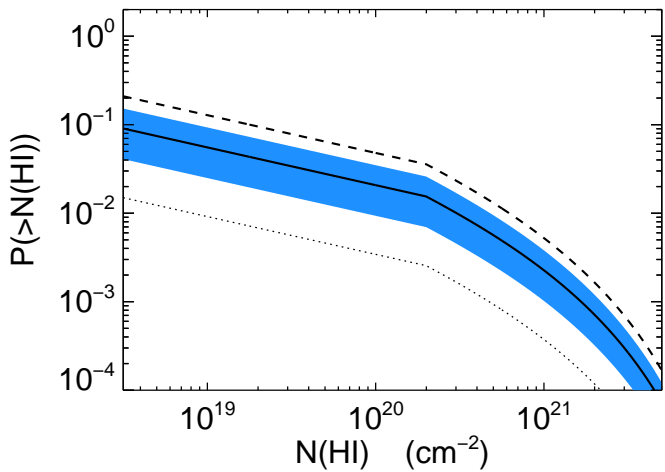}}}}
\caption{\label{clustering} 
Probability of coincident absorption as a function of column density
threshold. The lower dotted line shows the cosmic average probability for
finding an absorber within a $\pm$400 km/s window at redshift $z=2.7$ in the
absence of clustering.  The solid line represents the probability for
finding an absorber in sightline B, 13.8" away from and within $\pm$400 km/s of
an absorber detected in sightline A (or vice versa). This prediction assumes
that high-column density absorbers have a power law auto-correlation
function $\xi = (r\slash r_0)^(-\gamma)$ which is the same as the DLA-LGB
correlation function measured by Cooke et al. (2006) ($\gamma = 1.6; r_0 =
2.66^{+1.2}_{-1.3}$), and the blue-shaded region represents their
$\pm 1 \sigma$ errors. The dashed illustrates the effect of a steeper
correlation function ($\gamma = 2.1; r_0 = 2.8$) which was found to be
consistent with a subset of the Cooke et al. (2006) data.
}
\end{figure}

In the absence of clustering, the line density of absorption line
systems per unit redshift above the column density threshold N(HI) 
is given by the cosmic average
$\left\langle\frac{dN}{dz}\right\rangle(> $N(HI),$z)$. At an
average location in the universe, the probability of finding an
absorber in a background quasar spectrum within the redshift interval
$\Delta z = 2(1+z)\Delta v\slash c$, corresponding to a velocity
interval $\pm\Delta v$ is simply $P=\langle dN\slash dz \rangle \Delta
z$.  For a close pair of quasar sightlines, the presence of an
absorber in one of the sightlines implies an increased probability of
finding an absorber at a similar redshift in the neighbouring
sightline.  If the quasar sightlines have a transverse separation $R$,
and assuming that one searches a velocity interval $\pm \Delta v$
about the absorber redshift, we follow Hennawi \& Prochaska (2007) and
express the enhanced line density in the neighbouring sightline,
$\frac{dN}{dz}(R,\Delta v)$, in terms of a transverse correlation
function $\chi_{\rm \perp}(R)$ as
 
\begin{equation} \frac{dN}{dz}(R,\Delta v) =
\left\langle\frac{dN}{dz}\right\rangle \left[1 + \chi_{\perp}(R,\Delta
  v)\right]
\label{eqn:clust}
\end{equation}

where $\chi_{\rm \perp}(R,\Delta v)$ is given by an average of the 3-d
absorber auto-correlation function, $\xi(r)$, over a cylindrical
volume with cross-section $A$ equal to the absorption cross section,
and height given by the length in the line-of-sight direction
corresponding to the velocity interval $2\Delta v$. Provided that we
are in the `far-field' limit, where the transverse separation is much
larger than the absorber cross section $R \gg {\sqrt A}$, it is a very
good approximation to replace the volume integral over the cylinder
with a line integral over the range $L=2\Delta v\slash a H(z)$, where
  $a$ is the scale factor and $H(z)$ is the Hubble constant at
  redshift $z$ (see Hennawi \& Prochaska 2007 for details).

Studies of the clustering of high-column density absorbers, such as DLAs, are
hampered by the relatively small number of absorber pairs known, which
is in turn limited by the number of quasar pairs with separations
smaller than the absorber auto-correlation length. These poor
statistics can be circumvented if instead one cross-correlates
absorbers with galaxies. Cooke et al. (2006) recently measured the
clustering of LBGs around DLAs at $z \sim 3$ and measured a best fit
cross-correlation length of $r_0=2.66^{+1.2}_{-1.3}~\hMpc$ assuming a
fixed power-law slope of $\gamma=1.6$ (see also Adelberger et
al. 2003; Gawiser et al. 2001; Bouch\'e \& Lowenthal 2004). This
clustering measurement is very close to the auto-correlation of LBGs,
indicating that the auto-correlation of DLAs is similar in strength.

In Figure~\ref{clustering} we use the Cooke et al. (2006) measurement and
equation~\ref{eqn:clust} to estimate the probability of detecting
coincident high-column density absorbers as a function of column
density threshold. The lower dotted line shows the cosmic average
probability $P=\langle dN\slash dz \rangle \Delta z$ for finding a single
absorber within a $\pm$400 km/s window at redshift $z=2.7$ in the absence
of clustering. Calculating this probability requires an expression for
the column density distribution $f(N)$, which we obtained by joining
the the $f(N)$ for DLAs log N(HI) $\ge 20.3$ (Prochaska et
al. 2005) with that for sub-DLAs $19 < \log $ N(HI)$ < 20.3$ (O'Meara et
al. 2007). The solid line represents the probability for finding an
absorber above a given N(HI) in a second sightline, given the presence
of an absorber in the first sightline within a $\pm$400 km/s 
velocity window.  This calculation uses the Cooke et
al. (2006) best-fit measurement for $\gamma = 1.6$ and is
performed for the specific case of a 13.8 arcsec separation.
The blue-shaded
region represents the range permitted by the $\pm 1 \sigma$
errors in $\gamma$ quoted by Cooke et al. (2006). 
For this model, the probability of finding coincident sub-DLA
absorption (both absorbers N(HI) $>$ 19.5) in the SDSS 1116+4118 AB binary
within a $\pm$400 km/s window is about 3\%.  In the absence of clustering,
this value is about 0.5\%, i.e. the random probability of finding
a sub-DLA in a $\pm$400 \kms\ velocity window based on the number density
distribution alone.

The smallest separation probed by the Cooke et al. (2006) measurement
is (co-moving) $\sim 400~$\hkpc, or about 40\% larger than the
separation of our quasar pair sightline. The auto-correlation
functions of high redshift galaxies tend to become progressively
steeper on (co-moving) scales less than about 
1~ $h^{-1}$ Mpc, characteristic of the
sizes of dark matter halos (Coil et al. 2006; Ouchi et al. 2005;
Lee et al. 2006; Conroy et al. 2005), and
similar behaviour might be expected around the absorbers considered
here. To illustrate the effect of a steeper correlation function, the
dashed line in Figure \ref{clustering} shows the probability for coincident
absorption for a correlation function with $\gamma = 2.11$ and
$r_0=2.81$. This choice was motivated by the fact that Cooke et
al. (2006) found that a subset of their data (11 of the 15 DLA
sightlines) were best-fit with cross-correlation function parameters
$\gamma = 2.11^{+1.3}_{-1.4}$ and $r_0=2.81^{1.4}_{_2.0}$. 
For this model, the probability of finding coincident sub-DLA
absorption (both absorbers N(HI) $>$ 19.5) in the SDSS 1116+4118 AB binary
within a 400 km/s window is about 8\%.

Although the probabilities that we determine from the clustering analysis
are not large ($<$ 10\% for a single coincident pair), they are 
more than an order of magnitude more likely than the probability of
absorption from a single large galaxy (previous sub-section).  
Previous observations of some wide separation QSO groupings have also
concluded that intervening (super-)clusters may be responsible for
correlated absorption (e.g. Jakobsen et al. 1986, 1988; Francis
\& Hewett 1993), although the scales probed here are an order
of magnitude smaller than those studies.  We note that
one of the coincident absorbers is separated from the systemic
redshift of the QSO by  $\sim$ 3000 and 4500 km/s ($z_{\rm abs}$ = 2.94
in QSO A/B respectively).  Such `proximate' DLAs (PDLAs) have been
shown to exhibit excess clustering around QSOs (Ellison et al.
2002; Russell, Ellison \& Benn 2006; Prochaska, Hennawi \& Herbert-Fort
2007).
The probability for coincident absorption close to the QSO
redshift may therefore be further enhanced beyond the calculation
above.

Finally, we note that these clustering statistics are not affected
by any \textit{a priori} knowledge of the presence of a DLA in either
sightline.  The analysis \textit{assumes} the presence of one DLA
and \textit{calculates} the probability that a matching DLA is found
in the second sightline.  Of course, all pairs with at least one
DLA (i.e. regardless of whether there is a coincident absorber
in the second sightline) should be included in order to calculate
unbiased statistics.  Currently, SDSS 1116+4118 AB is our only
fully analysed example, although we discuss in the final section
the prospects for improving these statistics.  

\subsection{Implications for line of sight observations}\label{LOS_sec}

\begin{figure}
\centerline{\rotatebox{0}{\resizebox{9cm}{!}
{\includegraphics{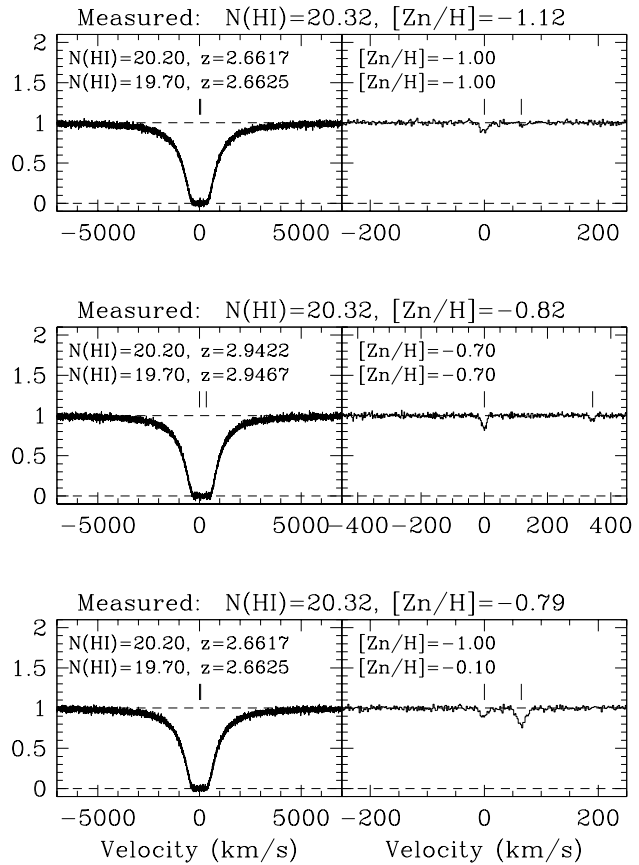}}}}
\caption{\label{blend} Illustration of the potential effect of
line of sight blending; \lya\ absorption is shown in the left-hand
panels and Zn~II $\lambda$ 2026 is shown in the right-hand panels.  
In all cases, the labels within the
panels show the input parameters of the two individual
simulated absorbers and the labels above the panels indicates
the value that would be measured for the combination of the two.
}
\end{figure}

Regardless of the physical interpretation of the coincident
absorption (single large galaxy, clustering etc.), the
correlated absorption has interesting ramifications for
single line of sight observations.    The most simplistic
extrapolation of the three intervening absorbers towards this one
binary system indicates that $\sim$ 2/3 of DLAs are not simple systems. 
From a different spatial
perspective, these spatially resolved systems could be superimposed
along a single sightline leading to an absorber that is a composite
of two (or more) components.  Whilst the consideration and treatment
of complex absorption systems consisting of many super-imposed `clouds'
is not new (e.g. Jenkins 1986; M\o ller, Jakobsen \& Perryman 1984), 
the data presented here
supply the first observational suggestion that superpositions of multiple
\textit{galaxies} may be relevent.  For the velocity differences observed
here (65 -- 340 km/s) the Ly$\alpha$ would appear as a single trough
with a column density indistinguishable from the sum of the components.
Indeed, only DLAs whose relative velocities are above $\sim$ 1000 km/s
would be recognised as multiple absorbers from the Ly$\alpha$ line
alone (depending on the column densities, since higher N(HI) systems
need to be separated by larger velocities for identification).
Although a two DLA blend could theoretically be identified
by the asymmetry between the blue and red wings of the profile,
in practice the delicate process of continuum fitting could
obscure this.  Difficulties include the complex underlying
form of the QSO continuum (such as emission lines), flux
suppression in the \lya\ forest and, in the case of cross-dispersed
spectra, complexity of removing the blaze, joining the orders
and poor flux calibration.   Indeed, when fitting the continuum 
around the Ly$\alpha$ trough of a DLA, curvature is often
required to yield a normalized spectra that can be well-fitted
with a damped profile (e.g. Prochaska et al. 2003), which
could lead to oversight of a blended absorber.

Cases of (partially) resolved multiple
DLAs have previously been reported by several authors (e.g.
Wolfe et al. 1986; Ellison \& Lopez 2001; Lopez \& Ellison 2003; 
Prochaska et al. 2003).  P\'eroux et al. (2003) have also noted
that 30\% of their sub-DLA sample are associated with another
absorber.
The impact of superimposed absorbers depends on a variety of factors,
such as relative velocities, N(HI) and metallicities.  We illustrate
three possible combinations in Figure \ref{blend}.  In the first scenario,
a DLA is blended with a sub-DLA, both of which have the same
metallicity, but due to the lower N(HI) of the sub-DLA, its metals
are below the detection threshold of the spectrum, leading to
an under-estimate of the abundance in the blended sightline
(top panel, Figure \ref{blend}).  
A similar under-estimate would result
if the velocity difference is large enough that the metal
lines from the second system are not included in the fit (middle
panel, Figure \ref{blend}).  In practice, however, this 
\textit{particular} situation
is unlikely to occur since normally other, stronger, metal lines
such as Fe~II would signal the existence of high velocity gas.
Alternatively, if a blended sub-DLA had a higher intrinsic abundance,
as has been proposed to be systematically the case (e.g. P\'eroux
et al. 2006)
we would infer an artificially high metallicity for the DLA (bottom
panel, Figure \ref{blend}).  Dessauges-Zavadsky et al. (2006)
have noted that the abundance ratios of different velocity
components in single line of sight DLAs are often non-uniform,
which is expected if the system is a blend of more than one galaxy.
These effects are of course most severe when the N(HI) values
of the blended absorbers are close in value; a sub-DLA with
log N(HI) = 19.0 will have little impact on a DLA with log N(HI) = 21.0.

The superposition of DLAs in single lines of sight may partially 
explain the often complex nature observed in metal
line profiles and the extreme rarity of DLAs with simple velocity
structures.  The potential for composite absorption by multiple
structures in a single line of underlines the difficulty of using
the kinematics of DLAs to infer their nature (Prochaska \& Wolfe
1997; Haehnelt, Steinmetz \& Rauch 1998).

\subsection{Does a maximum likelihood estimation of absorber
give an accurate DLA size?}

Given the clumpy distribution of DLA-column density gas in our simulated
galaxies, it is interesting to ask whether statistics from pairs of
QSOs can give an accurate picture of galaxy `size'.  We therefore
prepare a mock catalog of pairs of sightlines from which
we calculate a maximum
likelihood estimate of absorber size based on the coincidences and
anti-coincidences of DLA absorption (e.g. McGill 1990;
Dinshaw et al. 1997).  Whilst relatively
simple to implement, this technique assumes a simple geometry (usually
a spherical halo or cyclinder) and does not account for density fluctuations
and other small scale structure.  Nonetheless, this
technique has previously been used for C~IV and Mg~II absorbers
(Lopez, Hagen \& Reimers 2000; Ellison et al. 2004), but the statistics
are not yet available to attempt this for DLAs.
For pairs of sightlines intersecting spherical clouds, the probability that
a halo is intersected by the second line of sight, 
given that it is seen in the first, is given by

\begin{equation}
\label{phi}
\phi (X)=\frac{2}{\pi}\left\{ \arccos \left[ X(z)\right] - X(z)
\sqrt{1-X(z)^2} \right\}
\end{equation}

for $0 \le X(z) \le 1$ and zero otherwise.  Here, $X(z)=S(z)/2R$ where 
$S(z)$ is the line of sight separation and $R$ is the absorber radius.
In order to make maximum use of the information available
we actually want to calculate the probability that both lines of sight are
intersected, if $either$ line of sight shows absorption.  
This probability is given by

\begin{equation}
\psi=\frac{\phi}{2-\phi}
\end{equation}

\begin{figure}
\centerline{\rotatebox{270}{\resizebox{6cm}{!}
{\includegraphics{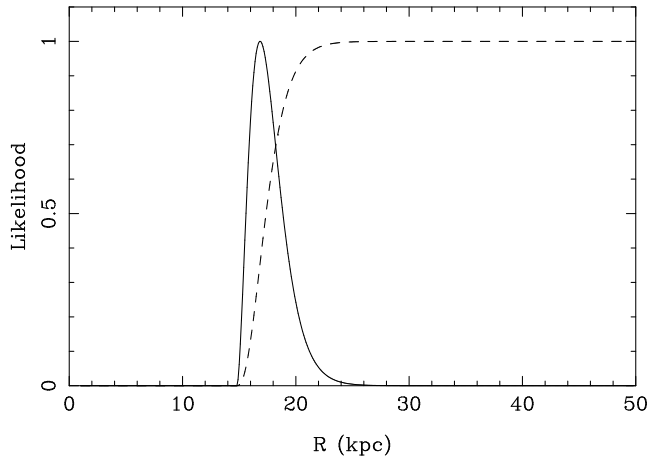}}}}
\caption{\label{max_like} Maximum likelihood distribution (solid
line) and cumulative distribution (dashed line) of the size of
DLA absorbing gas from a mock catalog of 50 pairs of sightlines
for galaxy K15 at $z=3.0$.  The most likely radius from this
simulated catalog is found to be 16.8 \hkpc\ with 95\% confidence
limits of 15.4 and 21.6 \hkpc.
}
\end{figure}

And the likelihood function as a function of radius is given by

\begin{equation}
\label{like}
{\cal L}(R)=  \prod_i \psi\left[ X(z_i)\right] \prod_j \left\{1-\psi\left[
X(z_j)\right] \right\},
\end{equation}

where $i$ and $j$ denote the number of coincidences and anti-coincidences.
The success of this technique relies on having a sufficient number
of coincidences as well as anti-coincidences to constrain the likelihood
function.  From a mock catalog of 50 (half each from the XZ and XY
orientations) pairs of sightlines with the
impact parameter of the first DLA absorber $<$ 30 kpc from the
centre of the galaxy and with all line of sight separations $<$
30 kpc, we have 5 coincidences and 45 anti-coincidences for galaxy
K15 at $z=3.0$ (middle left panels of Figures \ref{sim_box_xy_60}
and \ref{sim_box_xz_60}).
The results of the maximum likelihood analysis for this mock catalog
are shown in Figure \ref{max_like}; we determine a most likely radius
of 16.8$^{+4.8}_{-1.4}$ \hkpc\ (95\% confidence limits).  
A smaller sample, or wider pair separations
would have led to fewer coincidences and hence poorer constraints
on the absorber size. A visual comparison of this most likely size with the
spatial distribution of gas in Figure \ref{sim_box_xy_60}, shows that the maximum
likelihood method gives a quite reasonable estimate of the extent 
of DLA gas even when the distribution is clumpy; the 
covering fraction of gas with log
N(HI) $\ge$ 20.3 within 16.8 kpc is $\sim$ 50\%.  
Of course, a real catalog of pairs
will likely probe galaxies of very different sizes and will
therefore yield an average value.  For a smoother
morphology, such as the regular disk formed by the K33 galaxy at $z=2.3$
(see bottom right panel of Figure \ref{sim_box_xy_60}) we find
a most likely radius of $12.9^{+4.6}_{-2.6}$ kpc with 90\% of
the gas within this radius above the DLA N(HI) threshold (XY plane).

The compilation of a sample of 50 pairs with separations $<$ 30 kpc
($\sim 4$ arcsecs) in which at least one DLA is detected is a 
considerable challenge.
However, the required number of pairs is fewer if the separations 
are also smaller, because coincidences will be detected at a higher
rate and therefore constrain the absorber size better for a
smaller sample.  For example, if we repeat the maximum likelihood analysis
described above for the K15 galaxy at $z=3.0$, but with a maximum
pair separation of 20 kpc, from 4 coincidences and 16 anti-coincidences
we deduce a most likely radius of 13.2$^{+10.5}_{-2.5}$ kpc.
This means that if the simulations we have used in this work
are a reasonably accurate representation of the size of the DLA
cross-section in high redshift galaxies, close pairs with separations
$<$ 3 arcseconds or wide separation lenses will be best able
to observationally constrain DLA size.  

\section{Conclusions}

We have presented moderate resolution spectra of a pair of QSOs
at $z \sim 3$ separated by 13.8 arcseconds on the sky.  DLA or
sub-DLA absorption is identified at three intervening redshifts:
$z_{\rm abs} = 2.47, 2.66$ and 2.94.  For the two higher redshift
systems, absorption is seen in both lines of sight which have
proper transverse separations of $\sim$ 110 \hkpc.  
We measure chemical abundances for three of the five absorbers
detected in our ESI spectra and determine high metallicities
for two of them:  1/5 and 1/3 $Z_{\odot}$.  
Both absorbers also have quite high dust
depletions and for the $z_{\rm abs} = 2.94$ DLA we also find
that there is no $\alpha$ element enhancement, [S/Zn] $= +0.05$,
and measure a low fraction of molecular gas, log $f$(H$_2$) $< -5.5$
in this particular line of sight.

Although this is the first time
that transverse DLA absorption on this scale has been studied, if line of
sight blending occurs as often as the coincidences
observed in SDSS 1116+4118 AB, we demonstrate that this
could have important ramifications.
For example, the determination of chemical abundances can be affected
by several tenths of a dex and kinematics become challenging to
interpret.  Although more QSO pairs with separations
of a few tens to a few hundreds of kpc are clearly required to confirm
whether blending may really occur frequently in single line
of sight DLAs, our work identifies a potential complication in
the interpretation of DLA properties.

This is the first time that DLAs have been studied on a transverse 
scale $>$ 10 \hkpc\ and offers an opportunity to study their
size and environment.  By producing
artifical pairs of sightlines through high resolution galaxy
simulations, we conclude that the coincident absorption is unlikely
to be associated with a single large galaxy, or a galaxy plus
satellite system.  Instead, from a clustering analysis, 
we find that the statistically more
likely scenario is one in which the binary sightlines intersect
multiple galaxies.  However, the probability for coincident
sub-DLA absorption is still $<$ 10\%, so more observations
of QSO pairs are required to more robustly interpret our data.

The prospects for extending the analysis of absorption in
QSO pairs is promising, thanks to the large sky coverage of the SDSS.
Hennawi et al. (2006b) estimate that the current 8000 deg$^2$
of SDSS imaging contains $\sim$ 140 pairs with $\Delta \theta
< 25$ arcsecs and $z>1.8$.  Many of these will require follow-up
spectroscopy since the blue coverage of the SDSS spectra only reaches
down to $z_{\rm abs} \sim 2.2$.  However, based on absorber number densities,
Hennawi et al. (2006b) estimate that the SDSS sample of pairs would 
contain $\sim$ 70 sub-DLAs or DLAs.  Measuring the rate of
coincident absorption in a sample of this size may, at last,
place the scale of DLA absorption and the structure of its
environment on solid ground.

\section*{Acknowledgments}

The ESI spectra were reduced using software written by Jason
X. Prochaska, who was generous with his time and advice during the
data reduction process.  SLE acknowledges the hospitality of the Dark
Cosmology Centre in Copenhagen where some of this work was done. JFH
is supported by NASA through Hubble Fellowship grant \# 01172.01-A
awarded by the Space Telescope Science Institute, which is operated by
the Association of Universities for Research in Astronomy, Inc., for
NASA, under contract NAS 5-26555.   CLM is supported by the David and Lucile
Packard Foundation. We acknowledge useful discussions
with Alexei Razoumov and John O'Meara.  The TreeSPH
simulations were performed on the SGI Itanium II facility provided by
DCSC. The Dark Cosmology Centre is funded by the DNRF.
The data presented herein were
obtained at the W.M. Keck Observatory, which is operated as a
scientific partnership among the California Institute of Technology,
the University of California and the National Aeronautics and Space
Administration. The Observatory was made possible by the generous
financial support of the W.M. Keck Foundation.  The authors wish to
recognize and acknowledge the very significant cultural role and
reverence that the summit of Mauna Kea has always had within the
indigenous Hawaiian community.  We are most fortunate to have the
opportunity to conduct observations from this mountain.

\end{document}